\documentclass[12pt]{article}

\usepackage{graphicx}
\usepackage[space]{grffile}
\usepackage{latexsym}
\usepackage{textcomp}
\usepackage{longtable}
\usepackage{tabulary}
\usepackage{booktabs,array,multirow}
\usepackage{natbib}
\usepackage{url}
\usepackage{hyperref}
\hypersetup{colorlinks=false,pdfborder={0 0 0}}
\usepackage{etoolbox}
\usepackage{placeins}
\usepackage{amsmath,amssymb,amsfonts,bm}    
\usepackage{lineno,hyperref}
\usepackage{algorithm2e}
\usepackage{pdfpages}
\usepackage{subcaption}
\usepackage{placeins}
\usepackage{url}
\usepackage{xcolor}
\usepackage{natbib}
\usepackage{mathtools}
\usepackage{titlesec}
\usepackage{multirow}
\titlelabel{\thetitle.\quad}

\begin{document}

	\def\spacingset#1{\renewcommand{\baselinestretch}%
		{#1}\small\normalsize} \spacingset{1}
	
	\pagenumbering{arabic}


		\title{\bf Censored Regression for Modelling International Small Arms Trading and its "Forensic" Use for Exploring Unreported Trades}
		\author{Michael Lebacher\thanks{Department of Statistics, Ludwig-Maximilians-Universit\"at M\"unchen, michael.lebacher@stat.uni-muenchen.de}\hspace{.2cm}, Paul W. Thurner\thanks{Department of Political Science, Ludwig-Maximilians-Universit\"at M\"unchen}\hspace{.2cm}
			and 
			G\"oran Kauermann\thanks{Department of Statistics, Ludwig-Maximilians-Universit\"at M\"unchen}\hspace{.2cm}\thanks{The project was supported by the European Cooperation in Science and Technology [COST Action CA15109 (COSTNET)]. We also gratefully acknowledge funding provided by the German Research Foundation (DFG) for the project  KA 1188/10-1 and TH 697/9-1.}}
\date{}
		\maketitle

	\bigskip
	\begin{abstract}
		\noindent 	In this paper we use a censored regression model to investigate data on the international trade of small arms and ammunition (SAA) provided by the Norwegian Initiative on Small Arms Transfers (NISAT). Taking a network based view on the transfers, we not only rely on exogenous covariates but also estimate endogenous network effects. We apply a spatial autocorrelation (SAR) model with multiple weight matrices. The likelihood is maximized employing the Monte Carlo Expectation Maximization (MCEM) algorithm. Our approach reveals strong and stable endogenous network effects. Furthermore, we find evidence for a substantial path dependence as well as a close connection between exports of civilian and military small arms. The model is then used in a "forensic" manner to  analyse latent network structures and thereby to identify countries with higher or lower tendency to export or import than reflected in the data. The approach is also validated using a simulation study.

	\end{abstract}
	
	\noindent%
	\vfill
	
	\newpage
	\spacingset{1} 

\section{Introduction}
The Small Arms Survey Update 2018 indicates transfers of small arms in 2015 amounting to  5,7 billion (\citealp[p. 19]{SAA2018}) with a major share and highest increases in ammunitions (\citealp[p. 22]{SAA2018}). Given the often fatal consequences - civilian or military - of the availability of these arms for intrastate conflict and shootings as well as for interstate war, the absence of empirical evidence for supplier-recipient networks is surprising. A major reason behind this research gap are the notorious data deficiencies due to non-reporting and illicit trafficking (see \citealp[p. 29-46]{SAA2018}). Based on the only large-scale data base for small arms (\citealp{marsh2016})  we aim to analyse for the first time the small arms trading network. We integrate gravity models in a statistical network design to apply a forensic statistical analysis.

Starting with the seminal work of \citet{tinbergen1962}, the gravity equation was quickly established as a valuable tool of empirical trade research. 
The success of the model stems from its intuitive interpretation as well as its surprisingly strong empirical validity, see e.g.\ \citet{head2014}.  It is therefore not surprising that the concept was applied to all kinds of trade relations, including the international exchange of arms.
An early example for these applications is the work of  \citet{bergstrand1992}.  Although he doubted the suitability of the model for arms trade because of the strong political considerations in this area, the approach was taken up more recently. \citet{akerman2014} and \citet{Thurner2018} used the gravity model in order to explain whether Major Conventional Weapons (MCW) are exchanged. \citet{Inmaculada2017} rely on the framework of \citet{helpman2008} to investigate the influence of economic and political variables on the so-called extensive and intensive margin of MCW trade. The interplay between oil imports and arms exports is determined using a gravity model in \citet{bove2018}.
While the papers above focus on the exchange of MCW, in our paper we investigate transfers of small arms and ammunition (SAA) provided by the Norwegian Initiative on Small Arms Transfers (NISAT). This data is arguably even better suited for a gravity model since small arms are potentially less dependent on political decision making and many more trade occurrences are recorded.

We propose a network perspective on international SAA trade and conceptualize countries as nodes and transfers between them as directed, valued edges. Although gravity models are a standard tool for the analysis of dyadic data (\citealp{kolaczyk2009}), endogenous network effects are rarely incorporated in these models. We do so by connecting the idea of gravity models with the  spatial autoregressive (SAR) model adjusted to network data. 
Especially in sociology, SAR models are regularly used in a network context since the early eighties (\citealp{dow1982}; \citealp{doreian1984};  \citealp{doreian1989b}). They are  called {\sl network  autocorrelation models} in this strand of literature. More recently, network autocorrelation models became popular in political science applications, see for example \citet{franzese2007}, \citet{hays2010} and \citet{metz2017}. Here, it is assumed that actors with certain characteristics are embedded in a network and this embedding leads to contagion and/or spillover effects transmitted through the edges that relate the actors (\citealp{Leenders2002}). Hence, one presumes that the characteristics of actors are correlated because a specific social, political or economic mechanism is  connecting them. Note that the design of these models is different as compared to the usual set-up of gravity models since the outcome is related to the nodes, and the edges only represent indicators for node dependence. In this paper, we are interested in the dependencies among the \textsl{transfers} (instead of the actors) and account for outdegree, indegree, reciprocity and exogenous covariates. A similar model in a non-network context is the spatial gravity model (\citealp{lesage2008}), that accounts for spatial dependence of the exporter, the importer as well as for the spatial importer-exporter dependencies.

Contrary to the typical structure of trade data we observe a high degree of reported non-trade in SAA. In other words, the trade network has a large percentage of zero entries. To accommodate the zero inflation problem we employ a censored SAR model that can be  fitted using the Monte Carlo Expectation Maximization (MCEM) algorithm (\citealp{dempster1977}; \citealp{wei1990}). There are already several similar EM-based approaches that have been pursued. For instance \citet{suesse2017} use the EM algorithm in spatial econometric models, \citet{schumacher2017} apply an EM-based application to a censored regression model with autoregressive errors, and \citet{vaida2009} utilize EM estimation in a censored linear mixed effects model. In \citet{augugliaro2018l1} a similar estimation procedure is used in the context of fitting a graphical LASSO to genetic networks.

While the model application per se provides new insights into SAA trading, the ultimate  objective in this paper is to make use of the model to explore the validity of reported zero trades. This reflects a "forensic" objective, i.e.\ we estimate, whether unreported trades are likely to have happened based on the fitted model. Despite this idea is in line with forensic statistics and forensic economics (\citealp{aitken2004}; \citealp{Zitzewitz2012}) our goal is apparently less ambitious. We do not aim to provide statistical evidence that some states are under-reporting but we do want to investigate  \textsl{potential} under-reporting by utilizing the fitted network model.

This paper is organized as follows: after presenting the data in Section \ref{sec:description}, we explain the model and show how to proceed with estimation and inference in Section \ref{sec:model}. In Section \ref{sec:results} the results of the censored regression analysis are given and Section \ref{sec:reporting} provides the "forensic" analysis, accompanied by  a simulation study. Section \ref{sec:conc} concludes the paper.

\section{Data description} \label{sec:description}
Since 2001, the Geneva-based Small Arms Survey specializes on documenting the international flows of the respective products. However, only the Norwegian Initiative on Small Arms Transfers (NISAT, see \citealp{NISAT2018}) provides truly relational data necessary for applying network analysis. The NISAT database contains relational information on the trade of small arms, light weapons and ammunition (see also \citealp{marsh2016}). This information is collected from different sources as described in \citet{haug2002}. Although NISAT  represents the most reliable source of data regarding the exchange of small arms and light weapons, there is nevertheless an enormous amount of uncertainty inherent to arms trading data. This is especially true for light weapons where data quality and availability is partly very poor (\Citealp{LW2018}). Therefore, we restrict our analysis to small arms and the associated ammunition (SAA). See Table \ref{table:arms_included} in  Annex \ref{table:arms_included} for the types of small arms and ammunition included in the dataset. Note, that the NISAT database also contains data on sporting guns, which we excluded from the dataset since we are particularly interested in the export of small arms with potential military value. Actually, we will rely on transferred sporting guns volumes later as a  useful explanatory variable.
\begin{figure}[t!]
	\caption{Binary SAA trade network for the $59$ most relevant  countries in 1992 (left) and 2014 (right). Countries are indicated by grey nodes and transfers by edges in black.}
	\label{fig_netw}
	\centering
	\begin{subfigure}{6.5cm}
		\centering\includegraphics[trim={2.4cm 21.2cm 1.1cm 20.9cm},clip,width=6.8cm]{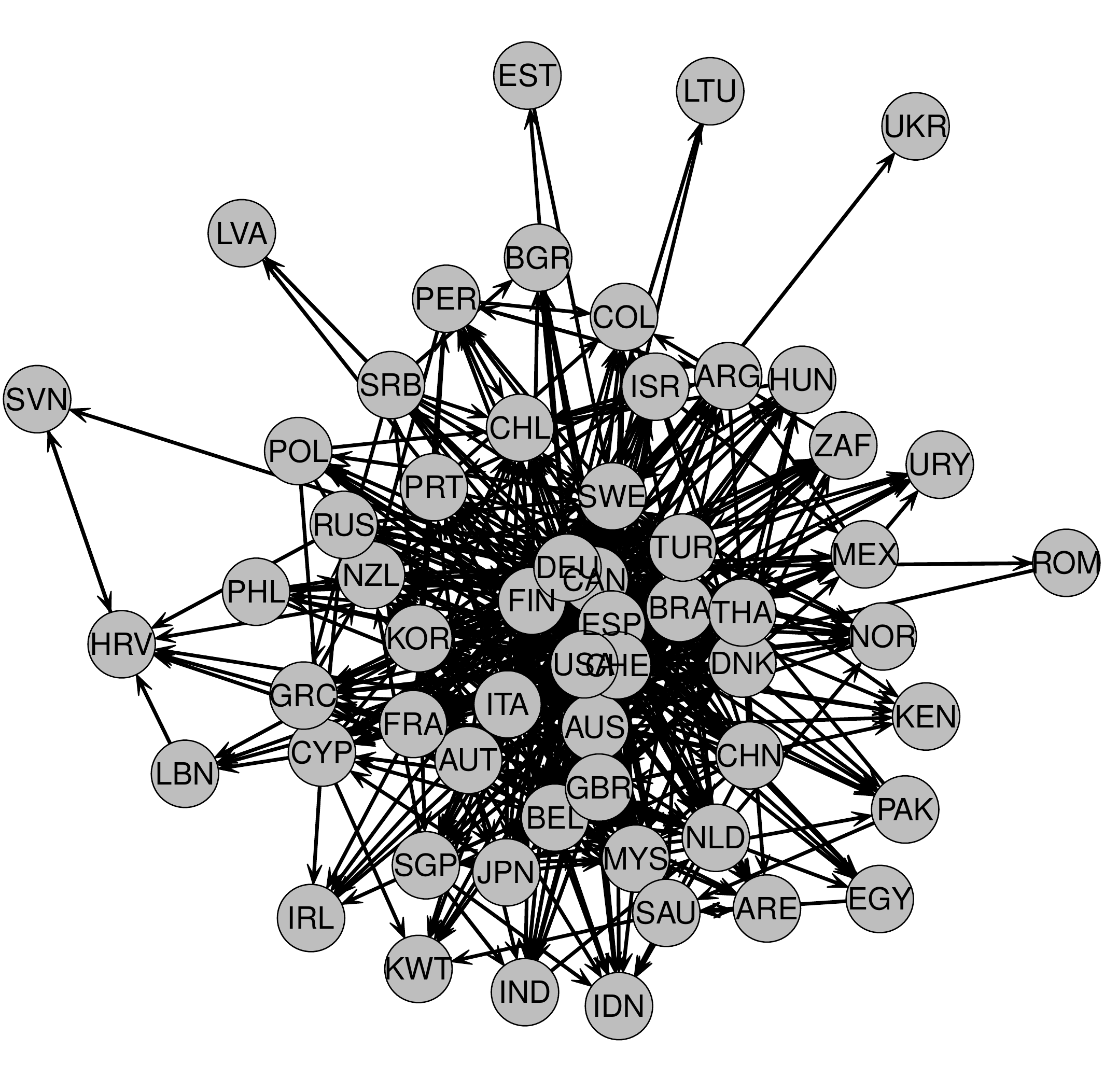}
	\end{subfigure}
	\begin{subfigure}{6.5cm}
		\centering\includegraphics[trim={2cm 21.2cm 1.5cm 20.9cm},clip,width=6.8cm]{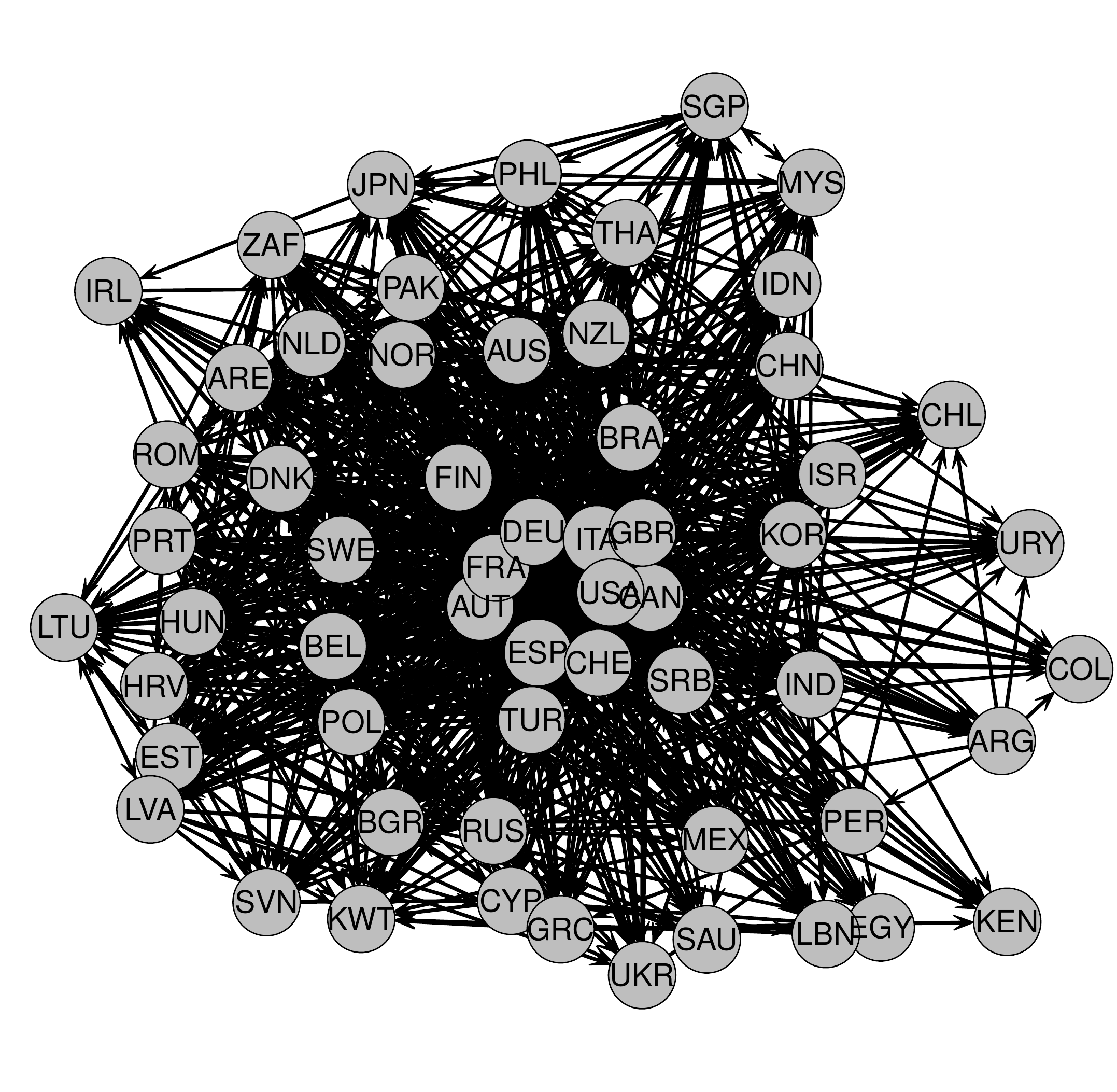}
	\end{subfigure}
	
	\centering
	\begin{subfigure}{6.5cm}
		\centering\includegraphics[trim={3cm 60cm 1.5cm 19.5cm},clip,width=6.8cm]{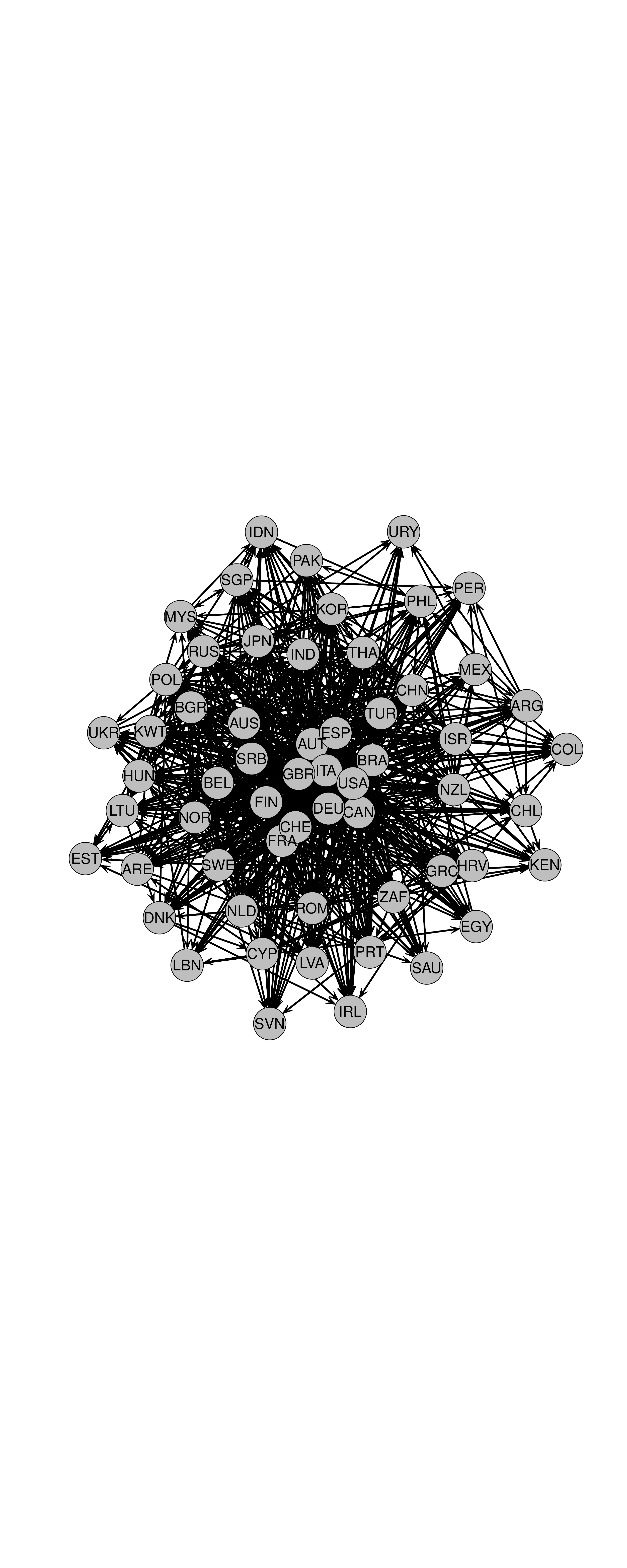}
	\end{subfigure}
	\begin{subfigure}{6.5cm}
		\centering\includegraphics[trim={2cm 60cm 1.5cm 19.5cm},clip,width=6.8cm]{networks_2014.pdf}
	\end{subfigure}
\end{figure}
In the remaining dataset, more than 86\,000 SAA transfers are recorded for the years 1992-2014, providing the exporting country, the importing country as well as the transferred arms category. The value of the export is measured in constant 2012 USD. In order to make estimation feasible, we restrict our analysis to a subnetwork and select those countries that account for the major share of the SAA trade activity. The resulting 59 countries (see Annex \ref{descrannex}, Table \ref{table:countries_included}) account for $73\%-91\%$ (depending on the year) of the total transfer volume and have participated in arms trade at least once in each year under study. Hence, we investigate the "core" of the international small arms trade network, balancing the trade-off between the number of countries included, the share of trade volume and the density of the subnetworks. In Figure \ref{fig_netw}, we show two binary networks for 1992 and 2014, with the countries represented as nodes and the arms transfers as directed edges among them. 

\begin{figure}[t!]\centering
	\caption{Aggregated exports (left) and density (right) in the SAA trade networks of the 59 most relevant countries. Countries with the highest export volume (Germany DEU, Italy ITA and United States USA) are highlighted.}	\label{features}
	\includegraphics[trim={0.1cm 0cm 0cm 0cm},clip,width=\textwidth]{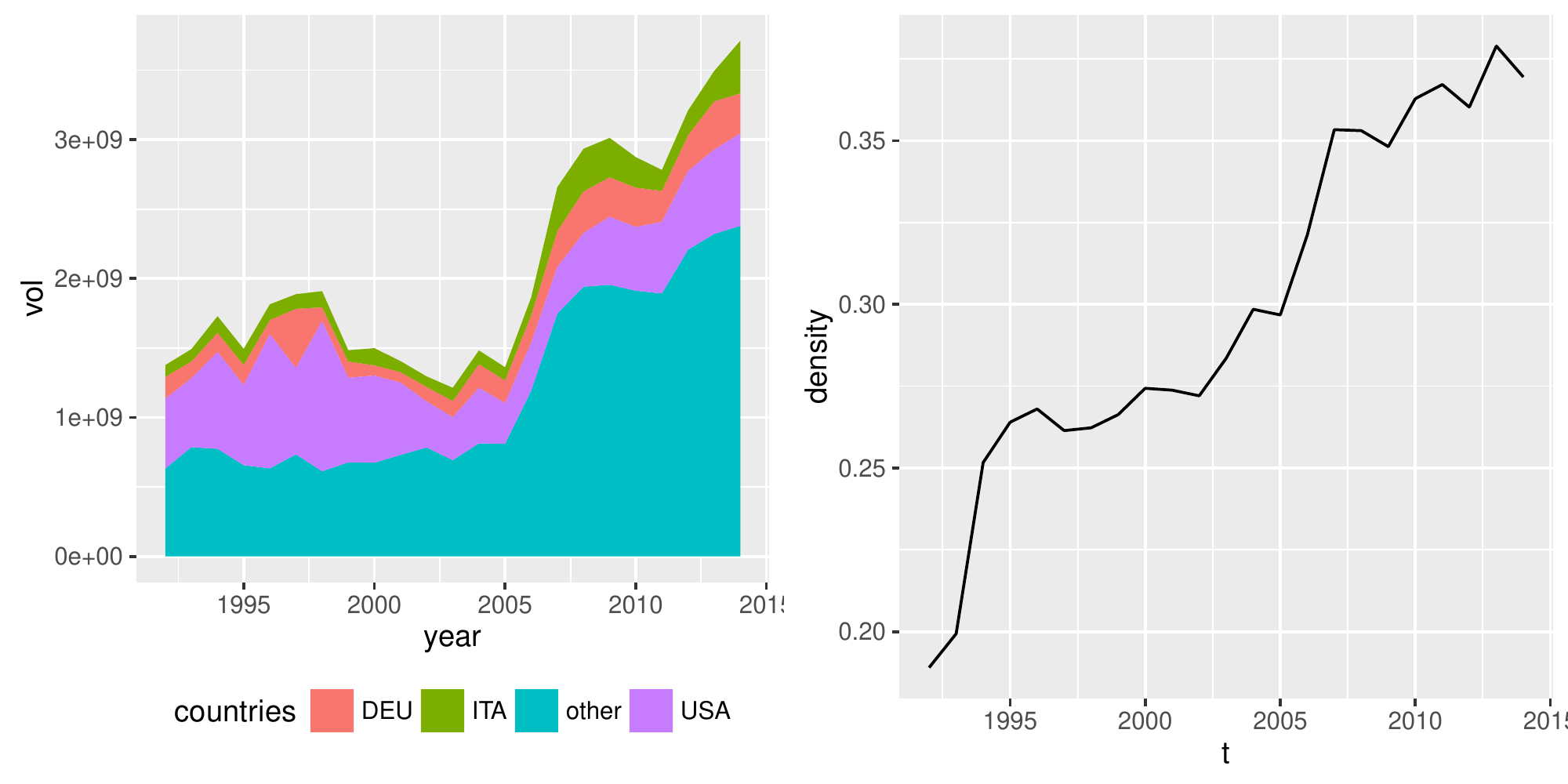}
\end{figure}

In the left panel of Figure \ref{features} we show the aggregated  exports for the most important exporters United States (USA), Germany (DEU) and Italy (ITA) together with the exports of the 56 other countries (other). On the right hand side of Figure \ref{features} we present the density, defined as the sum of existent edges divided by the number of potential edges.  Although the network can be without doubt described as a dense one (as compared to the density of social networks), the density is smaller than $0.2$ in the beginning and remains below $0.4$ in the subsequent recent years.

\FloatBarrier

\section{Regression model} \label{sec:model}
\subsection{General model}
Let $Y_t=(Y_{t,ij})\in \mathbb{R}^{n \times n}$ represent a network of transfers at the discrete time points $t=1,...,T$. At each time point $Y_t$ consists of $n$ nodes and $N=n(n-1)$ directed, continuous real valued edges, with diagonal elements $Y_{t,ii}$ left undefined. We set $\tilde{Y_t}=\text{vec}(Y_t)\in \mathbb{R}^N$ as the row wise vectorization of $Y_t$, excluding the diagonal elements.  In the following, we suppress the time index $t$ for ease of the notation and assume (after some suitable transformation) that in each time point $\tilde{Y}$ follows the autoregressive network model
\begin{equation}
	\tilde{Y}= \sum_{k=1}^{q}\rho_k W_k \tilde{Y} +X\beta+\epsilon, \text{ $\epsilon \sim \mathcal{N}_N(0,\sigma^2I_N)$}
	\label{eq:model1}
\end{equation}
with $\beta$ being a $p$-dimensional parameter vector for the design matrix $X$. 
The matrices $W_k$ are row-normalized weight matrices representing linear endogenous network effects, with parameters $\rho_k$ as their strength. Model (\ref{eq:model1}) is usually known as spatial autoregressive (SAR) model and we refer to \citet{leSage2009} for a more detailed discussion and to \citet{lacombe2004} or \citet{lesage2008} for similar models with multiple weight matrices. Standard software implementations that allow for  a likelihood based estimation of the model are mostly restricted to  the special case with $q=1$, for example in the \texttt{R} package \texttt{spdep} (\citealp{bivand2013}; \citealp{bivand2015}). The package \texttt{tnam} by \citet{tnam} allows for multiple weight matrices but is based on pseudo-likelihood estimation and therefore valid only if the weight matrices exclusively apply to exogenous covariates. Another possibility to estimate similar models is given by the package \texttt{ARCensReg} \citep{schumacher2017}, initially designed to fit models with autoregressive errors. Because of the similar mathematical structure, the package could be used to fit models with spatially dependent errors known as Spatial Error Models (SEM). In the given case however, the network structure is assumed to influence the response directly which prevents us from using the package.

Model  (\ref{eq:model1}) can be rewritten as
\begin{equation*}
	\tilde{Y}=\bigg{(}\underbrace{I_N- \sum_{k=1}^{q}\rho_k W_k}_{\equiv A(\boldsymbol{\rho})} \bigg{)}^{-1}(X\beta +\epsilon)=\underbrace{(A(\boldsymbol{\rho}))^{-1}}_{\equiv B(\boldsymbol{\rho})}(X\beta +\epsilon)=B(\boldsymbol{\rho})(X\beta +\epsilon),
\end{equation*}
where  the dependence on the $q$-dimensional parameter vector $\boldsymbol{\rho}=(\rho_1,...,\rho_q)^\mathrm{T}$ is made explicit for notational clarity. Similar as in \citet{besag1974} and given that all $N$ edges in the network are observed, their distribution is given by
\begin{equation}
	\label{eq:distribution}
	\begin{split}
		P(\tilde{Y}|X,\theta)&
		=\frac{1}{(2\pi \sigma^2)^{\frac{N}{2}}}|A(\boldsymbol{\rho})|\exp\bigg{\{}-\frac{(A(\boldsymbol{\rho})\tilde{Y}-X\beta)^\mathrm{T}(A(\boldsymbol{\rho})\tilde{Y}-X\beta)}{2\sigma^2}\bigg{\}}.
	\end{split}
\end{equation}
The parameter space of this model is restricted such that $A(\boldsymbol{\rho})$ is non-singular, which is ensured if the eigenvalues of $A(\boldsymbol{\rho})$ are real valued and greater than zero. 
\subsection{Censored regression model}

The above model is not directly applicable to our data since a large proportion of the SAA trade values is zero, expressing no (reported) SAA trade between countries $i$ and $j$. We therefore adapt model (\ref{eq:model1}) towards a utility model with censored observations. For each potential transfer from $i$ to $j$ there exists a utility of the transfer. This utility, however, only materializes in a transfer if it is higher than a certain threshold. Therefore, we assume that the probability model (\ref{eq:distribution}) applies to a network of partly observed latent variables, say $Z=(Z_{ij})$.  The relation among $Y_{ij}$ and $Z_{ij}$ is given by $Y_{ij}=\max(c,Z_{ij})$ for  $i,j=1,...,n$ and $c$ some threshold. Accordingly, we now set $\tilde{Y}=\text{vec}(Z)$ and label the observed utility $\tilde{Y}_o\in \mathbb{R}^{N_o}$ and the $N_m$ unobserved ones as $\tilde{Y}_m$. A reordering according to the observational pattern of $Y$ gives
\begin{equation*}
	\tilde{Y}=\left(
	\begin{array}{c}
		\tilde{Y}_o\\
		\tilde{Y}_m\\
	\end{array}
	\right) \sim \mathcal{N}_N \left( \left(
	\begin{array}{c}
		\mu_o\\
		\mu_m\\
	\end{array}
	\right) ,\left(
	\begin{array}{cc}
		\Sigma_{oo}&\Sigma_{om}\\
		\Sigma_{mo}&\Sigma_{mm}\\
	\end{array}
	\right)\right),
\end{equation*}
where $\tilde{Y}_m<c$ and $N=N_o+N_m$. Since the density of the network (see Figure \ref{features}) is roughly between $0.2$ and $0.4$ in all years, $N_m$ is always substantially larger than $N_o$.
The mean-covariance structure is given by
\begin{equation*}
	B(\boldsymbol{\rho})X\beta=\left(
	\begin{array}{c}
		\mu_o\\
		\mu_m\\
	\end{array}
	\right)\text{, }
	B(\boldsymbol{\rho})(B(\boldsymbol{\rho}))^\mathrm{T}\sigma^2=\left(
	\begin{array}{cc}
		\Sigma_{oo}&\Sigma_{om}\\
		\Sigma_{mo}&\Sigma_{mm}\\
	\end{array}
	\right).
\end{equation*}
In the following, we will denote all reordered matrices in the notation with double subscripts, i.e.\ $A_{oo}$ refers to the submatrix of $A$ where only interactions of observed variables $\tilde{Y}_o$ enter. 

\subsection{Monte Carlo EM estimation}
In order to estimate the unknown parameter vector $\theta=(\boldsymbol{\rho},\beta,\sigma^2) \in \mathbb{R}^{q+p+1}$, we employ the EM algorithm (\citealp{dempster1977}).
The {\sl complete log-likelihood}  $\ell_{comp}(\theta)$ is simply derived from  (\ref{eq:distribution}). We are interested in maximizing the {\sl observed log-likelihood} $\ell_{obs}(\theta)= \ell_{ \tilde{Y}_o}(\theta)+\ell_{\tilde{Y}_{m}|\tilde{Y}_o}(\theta)$, where the first part
is simply the multivariate normal density of the observed transfers. The second part equals
\begin{equation*}\footnotesize
	\begin{split}
		\ell_{\tilde{Y}_{m}|\tilde{Y}_o}(\theta)&=\log\bigg{(} \int_{(-\infty,c]^{N_m}}\frac{1}{\sqrt{(2\pi)^{N_m}|\Sigma_{m|o}|}}\exp\bigg{\{}-\frac{(U-\mu_{m|o})^\mathrm{T}\Sigma_{m|o}^{-1}(U-\mu_{m|o})}{2}\bigg{\}}dU \bigg{)},
		\label{censored}
	\end{split}\normalsize
\end{equation*}
where $\mu_{m|o}$ and $\Sigma_{m|o}$ are the first and second conditional moments.
Because $N_m$ is greater than $2\, 000$ in each year, the observed log-likelihood is numerically hard to evaluate (and even more so to maximize) with state of the art software implementation. As a solution, we apply the EM algorithm and maximize $ Q(\theta|\theta_0) \coloneqq \mathbb{E}_{\theta_0}[\ell_{comp}(\theta)|\tilde{Y}_o,X,\mathcal{M}] $ iteratively. The observation space is given by 
\begin{equation}
	\label{eq:M}
	\mathcal{M}=\{\tilde{Y}_m: \tilde{Y}_{m,1}<c,...,\tilde{Y}_{m,N_m}<c\}.
\end{equation}
\subsection{E-Step}
The E-Step essentially boils down to calculating the first two moments of a multivariate normally distributed variable $\tilde{Y}^c$
\begin{equation}
	\label{cdf}
	\tilde{Y}^c \sim \mathcal{N}_{N_m}(\mu_m+\Sigma_{mo}\Sigma_{oo}^{-1}(\tilde{Y}_o-\mu_o), \Sigma_{mm}-\Sigma_{mo}\Sigma_{oo}^{-1}\Sigma_{om})
\end{equation}
with restriction $\mathcal{M}$ from (\ref{eq:M}) applied to $\tilde{Y}^c$. Let those truncated moments be   $\mu_{m|o}^c$ and $\Sigma_{m|o}^c$ and define
\begin{equation}
	\label{condEx}
	\tilde{Y}^*=
	\left(
	\begin{array}{c}
		\tilde{Y}_o\\
		\mu^c_{m|o}\\
	\end{array}
	\right)
\end{equation}
as the vector that contains the observed values as well as the conditional expectation of the non-observed ones. Given the two moments, we can calculate the conditional expectation of the quadratic form  (see \citealp{mathai1992}):
\begin{equation}
	\label{eq:quadratic}
	\begin{split}
		S^*(\boldsymbol{\rho})&=\mathbb{E}_{\theta_0}[\tilde{Y}^\mathrm{T}(A(\boldsymbol{\rho}))^\mathrm{T}A(\boldsymbol{\rho})\tilde{Y}|\tilde{Y}_o,X,\mathcal{M}]=\\&=\text{tr}\bigg{(} (A_{mm}(\boldsymbol{\rho}))^\mathrm{T}A_{mm}(\boldsymbol{\rho})\Sigma_{m|o}^c\bigg{)}+(\tilde{Y}^*)^\mathrm{T}(A(\boldsymbol{\rho}))^\mathrm{T}A(\boldsymbol{\rho})\tilde{Y}^*.
	\end{split}
\end{equation}
Then, the function to maximize in the M-step is given by
\begin{equation}
	\label{Q_function}
	\begin{split}
		Q(\theta|\theta_0)&= -\frac{N}{2}\log(2\pi \sigma^2)+\log(|A(\boldsymbol{\rho})|)-\frac{(S^*(\boldsymbol{\rho})-2\beta^\mathrm{T}X^\mathrm{T}A(\boldsymbol{\rho})\tilde{Y}^{*}+\beta^\mathrm{T}X^\mathrm{T}X\beta)}{2 \sigma^2}.
	\end{split}
\end{equation}
In order to find the first and second moment of a truncated multivariate normally distributed variable,  \citet{vaida2009} use the results of \citet{tallis1961} on the moment generating function to provide closed form expressions of the E-Step. This, however, is not practicable in our setting as (a) software implementations of a multivariate normal distribution function are overstrained by the high dimension of our problem (the standard package in \texttt{R}, \texttt{mvtnorm} by \citet{Genz2016} is not able to process dimensions higher than $1\,000$) and (b) as noted by \citet{schumacher2017}, even if the distribution function could be evaluated, the closed form solution is computationally very expensive which leads to infeasible convergence times in applications with a high number of non observed values. The same is true for the  direct calculation using the moment generating function implemented in \texttt{R} by \citet{wilhelm2012}.

A practicable alternative consists in using the Monte Carlo EM (MCEM) algorithm (\citealp{wei1990}) where intractable expectations are replaced by  sample based approximations. In our specific case we use the \texttt{R} package \texttt{TruncatedNormal} by \citet{botev2017} in order to draw from the truncated multivariate normal distribution. An alternative would be to enrich the E-Step with a stochastic approximation step (SAEM algorithm, see \citealp{schumacher2017} for a detailed description) which reduces the number of simulations needed and is very efficient if the M-step is faster then the E-Step. In our specific application, the computational bottleneck comes with the M-Step and simulations showed that the SAEM converges more slowly than the MCEM algorithm.
\subsubsection{M-Step}
It is numerically more efficient to reduce the log-likelihood  to a profile log-likelihood by first maximizing with respect to $\beta$ and $\sigma^2$ and then with respect to $\boldsymbol{\rho}$. Using the derivatives of ($\ref{Q_function}$) with respect to $\beta$ and  $\sigma^2$ and defining $\hat{\beta}(\boldsymbol{\rho})$ and $\hat{\sigma}^2(\boldsymbol{\rho})$ as the solutions of the score equations as functions of $\boldsymbol{\rho}$  it follows that
\begin{equation}
	\begin{split}
		\hat{\beta}(\boldsymbol{\rho})&=(X^\mathrm{T}X)^{-1}X^\mathrm{T}A(\boldsymbol{\rho})\tilde{Y}^*\\ \hat{\sigma}^2(\boldsymbol{\rho})&=\frac{S^*(\boldsymbol{\rho}) - \tilde{Y}^{*\mathrm{T}}(A(\boldsymbol{\rho}))^\mathrm{T}H A(\boldsymbol{\rho})\tilde{Y}^*}{N},
	\end{split}
\end{equation}
where $H=X(X^\mathrm{T}X)^{-1}X^\mathrm{T}$ is the hat matrix. With $\kappa$ being a constant we can write the profiled function $\tilde{Q}(\cdot)$ as
\begin{equation}
	\label{eq:profile}
	\tilde{Q}(\boldsymbol{\rho}|\theta_0)=\kappa + \log(|A(\boldsymbol{\rho})|)-\frac{N}{2}\log\bigg{(}S^*(\boldsymbol{\rho}) - \tilde{Y}^{*\mathrm{T}}(A(\boldsymbol{\rho}))^\mathrm{T}H A(\boldsymbol{\rho})\tilde{Y}^*\bigg{)}.
\end{equation}
The expressions $(A(\boldsymbol{\rho}))^\mathrm{T}A(\boldsymbol{\rho})$ and $(A(\boldsymbol{\rho}))^\mathrm{T}HA(\boldsymbol{\rho})$  
have  derivatives
\begin{equation*}
	\begin{split}
		\frac{\partial (A(\boldsymbol{\rho}))^\mathrm{T}A(\boldsymbol{\rho})}{\partial\rho_k}&=- W_k-W_k^\mathrm{T}+2\rho_kW_k^\mathrm{T}W_k+\sum_{l\neq k}\rho_l(W_k^\mathrm{T}W_l+W_l^\mathrm{T}W_k)=:R_k(\boldsymbol{\rho})\\
		\frac{\partial (A(\boldsymbol{\rho}))^\mathrm{T}HA(\boldsymbol{\rho})}{\partial\rho_k}&=- HW_k-W_k^\mathrm{T}H+2\rho_kW_k^\mathrm{T}HW_k+\sum_{l\neq k}\rho_l(W_k^\mathrm{T}HW_l+W_l^\mathrm{T}HW_k)=:H_k(\boldsymbol{\rho}).
	\end{split}
\end{equation*}
Now define
\begin{equation*}
	R_k^*(\boldsymbol{\rho})=\text{tr}( R_{k,mm}(\boldsymbol{\rho})\Sigma_{m|o}^c)+(\tilde{Y}^*)^\mathrm{T}R_k(\boldsymbol{\rho})\tilde{Y}^*
\end{equation*}
which gives
\begin{equation}
	\label{eq:profile_gradient}
	\frac{\partial \tilde{Q}(\boldsymbol{\rho}|\theta_0)}{\partial \rho_k }=-\text{tr}(B(\boldsymbol{\rho})W_k)-\frac{N}{2}\frac{R_k^*(\boldsymbol{\rho})-\tilde{Y}^{*\mathrm{T}}H_k(\boldsymbol{\rho})\tilde{Y}^*}{S^*(\boldsymbol{\rho}) - \tilde{Y}^{*\mathrm{T}}(A(\boldsymbol{\rho}))^\mathrm{T}H A(\boldsymbol{\rho})\tilde{Y}^*}.
\end{equation}
Iteration between the E- and the M-step provide the final estimate $\hat{\theta}$. The variance of $\hat{\theta}$ can be calculated using \citet{louis1982} formula with more details on the practical implementation provided in the Supplementary Material.

\section{Application to the data}\label{sec:results}

\subsection{Covariates}
Considering model (\ref{eq:model1}) we need to specify the two major components of the model, namely (a) the covariates included in matrix $X$ and (b) the network related correlation structure.\\
{\sl Node Specific Variables:} Following standard applications (\citealp{ward2013}; \citealp{head2014}; \citealp{egger2016}; \citealp{Thurner2018}), we control for the logarithmic real GDP in constant 2010 USD  as a measure for the market size of the exporting and importing country. The data are provided by the  \citet{GDP2017}. For the two years 1993-1994 no reliable GDP data are available for Serbia, Croatia, Estonia, Latvia, Lithuania and Slovenia, we therefore assume that the GDP remained constant in the first three years for this countries.
In order to control for the potential influence of intrastate conflicts we insert a binary variable that is one if there is an intrastate conflict in the receiving country in the respective year and zero otherwise. The corresponding data is available from the webpage of the Uppsala Conflict Data Program (\citealp{ucdp2019}).

{\sl Edge Specific Distance Measures:}
Because of the strong empirical evidence that geographic distance is a relevant factor in trade (\citealp{disdier2008}), we control for the logarithmic distance between capital cities in kilometres (\citealp{gleditsch2013d}). In recent applications of the gravity model to arms trade (\citealp{akerman2014}; \citealp{Inmaculada2017}; \citealp{bove2018};  \citealp{Thurner2018}) it is argued that political distance measures in terms of regime dissimilarity must also be inserted in the gravity equation. We use the absolute difference of the polity IV index (\citealp{Polity2016}) between two countries, ranging from 20 (highest ideological distance) to zero (no ideological distance). Additionally, we include a dummy variable for formal alliances between the exporting and importing country, being one if the two countries have a formal alliance. The data is available from  \citet{Defagr2016} until 2012 and we assume that the alliances stay constant for the years 2013 and 2014. 

{\sl Edge Specific Trade Measures:} We control for lagged logarithmic SAA transfers by smoothing the past observed trade volume using a five-year moving average. In the years with less then five lagged periods available, the moving average is shortened accordingly. We call this path dependency, leading to inertias that arises because of diminishing transaction costs, trust relations, security aspects and potentially interoperability, and is a very important determinant in the MCW trade network (\citealp{Thurner2018}). 

Additionally, we enrich the model with a five year moving average of logarithmic civilian weapon transfers. The intuition behind that is that exports of SAA for military usage and civilian usage might be correlated. This is plausible because countries that export massive amounts of civilian arms also have the capabilities to produce military arms. The data is also provided by NISAT (\citealp{marsh2016}). Furthermore it seems plausible that there is a connection between the volume of small arms traded and the volume of  MCW. MCW transfers are recorded by the Stockholm International Peace Research Institute (SIPRI) and measured in so called trend indicator values (TIV). This measure represents the military value and the production costs of the transferred products. For detailed explanation of the data and the TIV see \citet{siprimeth2017, sipridata2017} and \citet{holtom2012}. We use a dummy variable that is one if there was an MCW transfer from country $i$ to $j$ in the actual year or in the four preceding years, zero otherwise. Additionally, we use the logarithmic sum of the exported TIV volumes in the actual year and the four preceding ones.

\begin{figure}[t]\centering
	\caption{Schematic representation of linear network effects. The focal edge in dashed grey.}	\label{figure:toy}
	\includegraphics[trim={0cm 12.55cm 7cm 2cm},clip,width=0.9\textwidth]{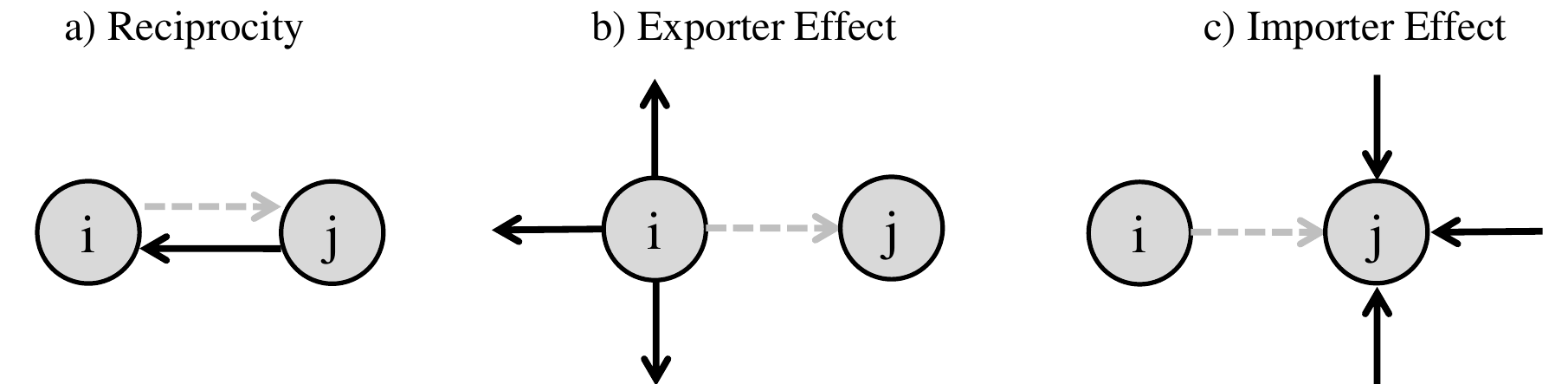}
	
\end{figure}

\subsection{Network structure} Next we specify the network specific effects represented by matrices $W_k$ in model (\ref{eq:model1}). We include three effects which are explained subsequently and visualized in Figure \ref{figure:toy}.

\textsl{Reciprocity:}  The reciprocity effect measures whether the export volume from country $i$ to country $j$ increases in the export volume from $j$ to $i$. In the given context it is a plausible assumption that countries tend to specialize in certain types of small arms and/or ammunition and therefore complement each other with their products. Mutual trade is likely to be encouraged by  political partnerships and indicates strategic elements, induced by bilateral agreements. The measure is also investigated in the context of commercial trade (e.g.\ \citealp{garlaschelli2005}; \citealp{barigozzi2010};\citealp{ward2013}).
In the arms trade literature, reciprocity is specified by \citet{Thurner2018}, with the finding that this is rather unusual in the context of MCW.

\textsl{Exporter and Importer Effect:} The exporter and the importer effect have their analogies in binary networks and can be interpreted as the valued versions of the outdegree and the indegree. The coefficient of the exporter effect measures whether the transfers going out from a certain exporter $i$ are correlated. A positive effect indicates the presence of "super-exporters". Contrary, the importer effect measures whether the imports of a certain importer $j$ are related, with a positive effect indicating "super-importers". The degree structure is a crucial feature of the SAA network because a rather small number of countries accounts for the major share of the trade volume, while a small share of (potentially identical) importing countries accounts for a great amount of the import volume. 

Before fitting the model we apply the natural logarithm to the data. This is necessary, because in its raw form the data is strongly skewed with a long tail. In the Supplementary Material the distribution of the log-transformed response is investigated

Hence, if the original trade matrices are given by $Y_t=(Y_{t,ij}) $, the elements of $\tilde{Y}_t$ are given by $\tilde{Y}_{t,ij}=\log(Y_{t,ij})$ if $Y_{t,ij}>0$ and are not defined if $Y_{t,ij}=0$. Furthermore, we define $d_t=\min(\{Y_{t,ij}>0\})$ as the lowest strictly positive value in the network at year $t$ and set $c_t=\log(d_t)$. That is, the threshold $c_t$ is defined such that at a given time point $t$ all transfers below the smallest observed log-transformed transfer in that sample are censored. Utility below the threshold $c_t$ implies that no transfer was carried out or was not recorded. Furthermore, we allow for time-varying coefficients by estimating each time-period separately. This relaxes the unrealistic assumption of time-constant effects for more than $20$ years and reduces the computational effort. Given these specifications, the final model is now given by
\begin{equation*}
	\begin{split}
		\tilde{Y}_{t,ij}&=X_{t,ij}^\mathrm{T}\beta_t +\underbrace{ \rho_{t,1} \tilde{Y}_{t,ji}}_{\text{Reciprocity}}+ \underbrace{\rho_{t,2} \frac{1}{n-2}\sum_{u\neq j} \tilde{Y}_{t,iu}}_{\text{Exporter Effect}} + \underbrace{\rho_{t,3} \frac{1}{n-2}\sum_{u\neq i}\tilde{Y}_{t,uj}}_{\text{Importer Effect}}+\epsilon_{t,ij},\\\text{ }\epsilon_{t,ij} &\sim \mathcal{N}(0,\sigma^2)\text{ for }i,j=1,...,n \text{, }i\neq j\text{, }t=1993,...,2014, n=59 \text{ and } N=3\,422. 
	\end{split}
\end{equation*}

\subsection{Results: Coefficients}
\begin{figure}\centering
	\caption{Time-series of annually estimated regression coefficients. Shaded areas give $\pm$ 2 standard errors. Colouring according to p-values, green: $p<0.05$, yellow: $p<0.1$ and red: $p>0.1$.}	\label{figure:coef}
	\includegraphics[trim={0cm 0cm 0cm 0cm},clip,width=0.85\textwidth]{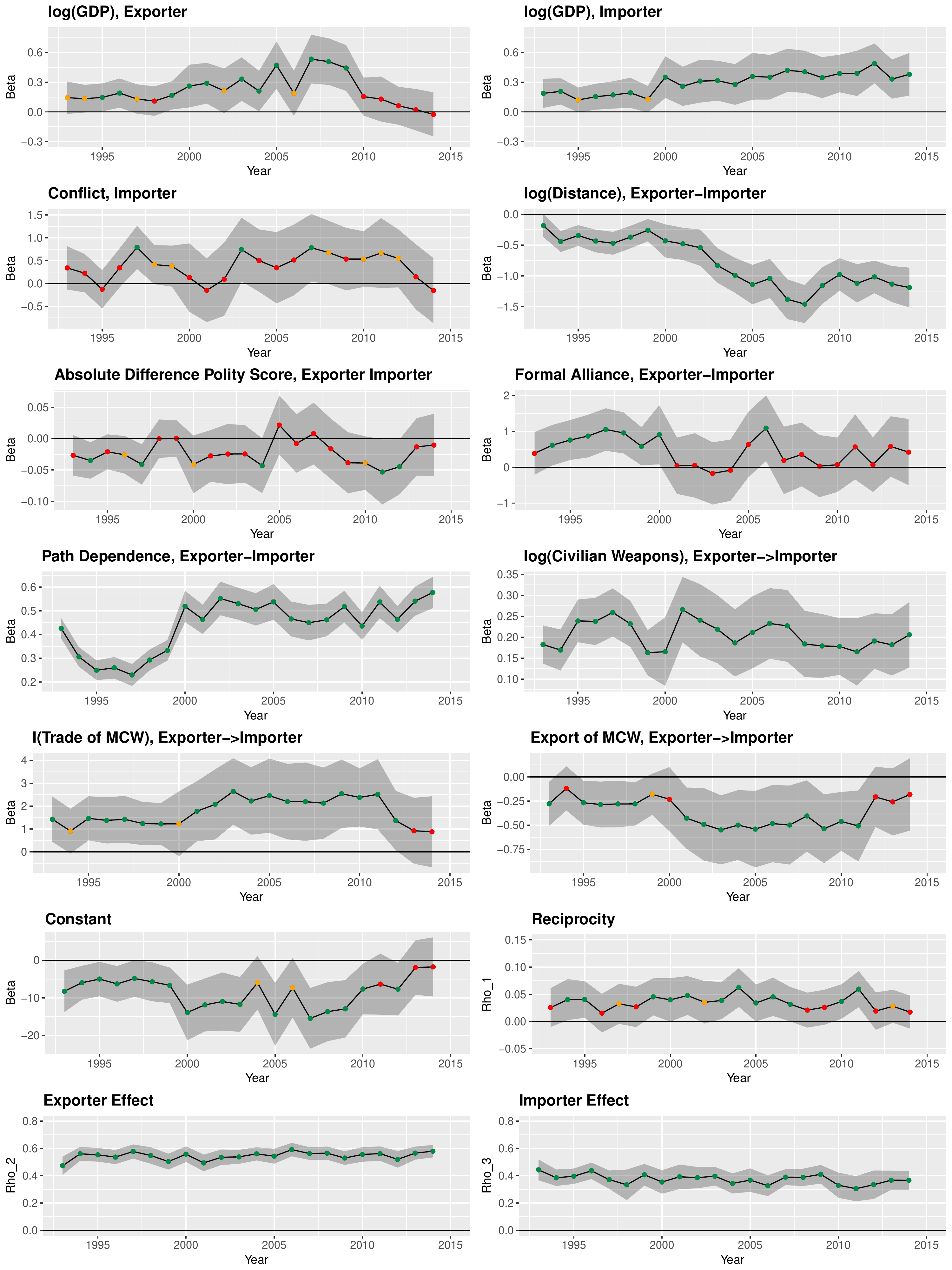}
	
\end{figure}
In Figure \ref{figure:coef} the time series of coefficients are plotted against time for the years 1993-2014. The shaded areas around the coefficients give two standard error bounds and the colouring of the point estimates reflects the respective significance level, the zero-line is depicted by solid black.

\noindent {\sl Covariates:} The exogenous covariates in the first row and in the second row on the right represent the standard gravity variables logarithmic GDP of the exporter and the importer as well as the geographic distance between them (second row, right panel). Overall, the expected results of the gravity equation hold, except for the logarithmic GDP of the exporter that tends to be insignificant and is close to zero in the most recent years. This is an interesting result, because it highlights the fact, that market size is not a prerequisite for producing and exporting internationally competitive SAA. This finding is in stark contrast to the insights on MCW by \citet{Thurner2018}. 

The strong negative effect of the geographic distance is significant in all years. Again, this is different as compared to MCW transfers where geopolitical strategy disregards distance.

Regarding the political security measures we find that the presence of a conflict in the importing country (second row, left panel) has a mostly positive but seldom significant effect while the coefficient on the dissimilarity of political regimes (third row, left panel) is mostly negative but also often insignificant. The coefficient on the dummy variable for formal alliances (third row, right panel) is positive in the beginning but almost permanently insignificant from 2001 on.

The large and consistently significant coefficients of the lagged moving average (fourth row, left panel) illustrates an important feature of the network, namely path inertia. Intensive transfer relationships in the past, strongly increase the export volume in the present. Similarly we find a strong connection between exporting civilian and military arms (fourth row, right panel). Looking at the relation between SAA and MCW trade we find that having traded MCW (fifth row, left panel) in the actual year or the in the four preceding ones has a strong positive effect -  at least until the last two years. However, at the same time the effect of the logarithmic sum of the TIV values (fifth row, right panel) has a negative and mostly significant effect. I.e.\ rather small transfers of MCW tend to coincide with SAA exports while dyads with high amounts of MCW exchange tend to transfer small arms to a relatively lower degree.

\noindent {\sl Network Structure:} On the right panel in the sixth row of Figure \ref{figure:coef} the coefficients for reciprocity are shown. The coefficients remain almost constant and positive with values between $0.02$ and $0.06$. As the coefficients often changes from significant to insignificant we infer that there is at least a tendency that mutuality increases the volume of arms exchanged.

The strongest endogenous effect is the exporter effect (bottom row, left panel) with coefficients that are consistently positive and significant. This indicates that the transfers stemming from the same exporter are indeed highly correlated and reflects the existence of "super-sellers" like the United States, Germany, Brazil or Italy. On the other hand, we also find a stable, positive and significant importer effect (bottom row, right panel). The fact that the two coefficients on the exporter and the importer effect are much higher than the reciprocity effect provides structural information about heterogeneity in the network. Being a strong exporter or sending to a strong importer increases the export volume more than simply having imported high amounts from the respective partner. 
\FloatBarrier
\section{"Forensic" statistical analysis}\label{sec:reporting}

\subsection{Under- and over-reporting}
Our model rests on the assumption that the SAA network is determined by a latent utility network  $Z_t$. Based on the joint distribution (\ref{eq:distribution}) we can in fact estimate the probability of $Z_{t,ij}$ being greater than the threshold $c_t$, given the covariates, the endogenous effects and the rest of the network.
In order to do so, let $Z_{t,-ij}$ represent the $(N-1)$-dimensional vector that contains the realized and the expected values of the latent variables, except the entry that corresponds to the transfer from $i$ to $j$. Because we are interested whether some latent transfers could have realized  according to the model, we form the expectations \textsl{without} the restriction that the latent transfers must be smaller than $c_t$.
Based on this, we define the conditional probability of a specific latent transfer being greater than the threshold by
\begin{equation*}
	\pi_{t,ij} = \mathbb{P}(Z_{t,ij}>c_t|X_{t,ij},Z_{t,-ij};\hat{\theta}_t).
\end{equation*}
By construction (see the Supplementary Material for the derivation), $\pi_{t,ij}$ is high for transfers that are observed in the dataset ($Y_{t,ij}>0$) and small for transfers that are not observed ($Y_{t,ij}=0$). However, we may calculate a high value of $\pi_{t,ij}$, i.e.\ a high probability for a realized transfer of arms, despite the data actually indicates $Y_{t,ij}=0$. We propose to consider this as potential \textsl{under-reporting}. Such a zero-record can happen due to random fluctuation, factors beyond the model as for example historical relationships, or because de-facto existent transfers have not been reported. Vice versa, we may obtain a low value of $\pi_{t,ij}$ although $Y_{t,ij}>0$. We label this  as \textsl{over-reporting}. This label is not intended to suggest that potentially over-reported transfers in fact never happened, but highlights transfers where our model attaches a lower level of latent utility than manifested in the data.  Naturally, our main "forensic" interest is in uncovering potential under-reporting.

Apparently, this requires the fixation of a threshold value for the probabilities. Based on Receiver-Operating Characteristic (ROC) curves, an optimal threshold value  $J_{t}$ can be found using Youden's $J$ statistic (\citealp{youden1950}). This value is optimal in the sense that it allows for a  separation such that both, sensitivity and specificity are maximized. This defines the binary network
\begin{equation*}
	\Pi_t=(I(\pi_{t,ij}>J_t)).
\end{equation*}
This network is now set into relation with the observed binary SAA trade
\begin{equation*}
	\Gamma_t=(I(Y_{t,ij}>0)).
\end{equation*} 
Comparing $\Pi_t$ and $\Gamma_t$, we can define the "forensic" network 
\begin{equation*}
	\Omega_t=(\omega_{t,ij})=\Pi_t-\Gamma_t
\end{equation*}
which in turn creates two new binary networks
\begin{equation*}
	\begin{split}
		\Omega_t^+=&(I(\omega_{t,ij}=1))\\
		\Omega_t^-=&(I(\omega_{t,ij}=-1)).
	\end{split}
\end{equation*}
For $\omega_{t,ij}=1$, the model predicted a transfer that is not present in the dataset, and for $\omega_{t,ij}=-1$, the model did not predict an actual transfer. Following our  convention from above we label $\Omega_t^+$ as the \textsl{under-reporting network} and  to   $\Omega_t^-$ as the \textsl{over-reporting network} of unpredicted but realized transfers.

\subsection{Simulation study of "forensic" power}\label{sec:sim}

Before we apply our model in a "forensic" matter to identify transfers with potential under-reporting we demonstrate the behavior of the model in a simulation study to explore its detection properties. We use two different settings in order to investigating how well the proposed approach identifies under-reporting. The first setting builds on  the following Data Generating Process (DGP1)
\begin{equation}\begin{split}
		\bm{\rho}&=(0.1,0.2,0.3)^\mathrm{T}\text{, }  \beta=(1,2,3,4,5)^\mathrm{T}\text{, }p=5\text{, }n=20\text{, }N=380\\
		X&\sim \mathcal{N}_{p}(\bm{1},I_{p})\\
		Z&\sim \mathcal{N}_N(B(\bm{\rho})X\beta,B(\bm{\rho})B(\bm{\rho})^\mathrm{T})\\
		\tilde{Z}_{ij}&=I(Z_{ij}>q_{0.75}(Z))Z_{ij}\text{, for }i \neq j=1,...,n.
	\end{split}
\end{equation}
Here, $q_{0.75}(Z)$ denotes the 75\% quantile and we are censoring the network towards an observed density of $0.25$. Note, that DGP1 is not subject to under-reporting and all censored responses are in fact below the censoring threshold.
The results of running DGP1  $100$ times and applying the estimation procedure are summarized in the Supplementary Material, indicating that the expected values approximate the latent variables very well and that we are able to find unbiased estimates despite the enormous amount of censoring.

In order to validate the forensic power of the model, we run a second experiment (DGP2), being a modified version of DGP1. To be precise, we are censoring again  75\% of the observations but only 65\% correspond to the lowest ones, while the remaining 10\% are randomly selected among the observations that are in fact \textsl{higher} than the threshold $q_{0.65}(Y)$. This share of observations represents the under-reporting. Again we run DGP2 100 times.
\begin{table*}
	\centering
	\captionsetup[subtable]{position = below}
	\captionsetup[table]{position=top}
	\caption{Schematic representation of the evaluation scheme used in the simulation study. True conditions in the columns and Estimated in the rows. \textbf{UR} denotes under-reporting and $\overline{\textbf{UR}}$ denotes censored observations that are not under-reported. Further abbreviations: True positive (TP), false positive (FP), false negative (FN) and true negative (TN). }
	\label{tab:scheme}
	\begin{subtable}{0.49\textwidth} 
		\centering
		\begin{tabular}{c|c|c|c||c}
			\multicolumn{5}{c}{TRUE}\\ \hline
			\multirow{4}{*}{Estimated}	& & \textbf{UR}&$\overline{\textbf{UR}}$& $\sum$ \\ \hline
			&		\textbf{UR} & 0 &FP&FP\\ 
			&		$\overline{\textbf{UR}}$ &0&TN&TN \\ \hline \hline
			&		 $\sum$& 0&$0.75N$&$0.75N$ \\ \hline
		\end{tabular}
		\caption{Classifier evaluation DGP1.}
	\end{subtable}%
	\begin{subtable}{0.49\textwidth} 
		\centering
		\begin{tabular}{c|c|c|c||c}
			\multicolumn{5}{c}{TRUE}\\ \hline
			\multirow{4}{*}{Estimated}	& & \textbf{UR}&$\overline{\textbf{UR}}$& $\sum$ \\ \hline
			&		\textbf{UR} & TP &FP&TP+FP\\ 
			&		$\overline{\textbf{UR}}$ &FN&TN&FN+TN \\ \hline \hline
			&		 $\sum$& $0.1N$&$0.65N$&$0.75N$ \\ \hline
		\end{tabular}
		\caption{Classifier evaluation DGP2.}
		
	\end{subtable}%
	\hspace*{4em}
\end{table*}

\begin{figure}[t!]\centering
	\caption{Results of DGP1 and DGP2. The top panel shows boxplots for the false positive rate (FPR) in DGP1 (left) and DGP2 (right). On the bottom boxplots for the true positive rate (TPR) and the false discovery rate (FDR) are provided for DGP2.}	\label{fig:eval}
	\includegraphics[trim={0.1cm 0.2cm 0cm 1cm},clip,width=\textwidth]{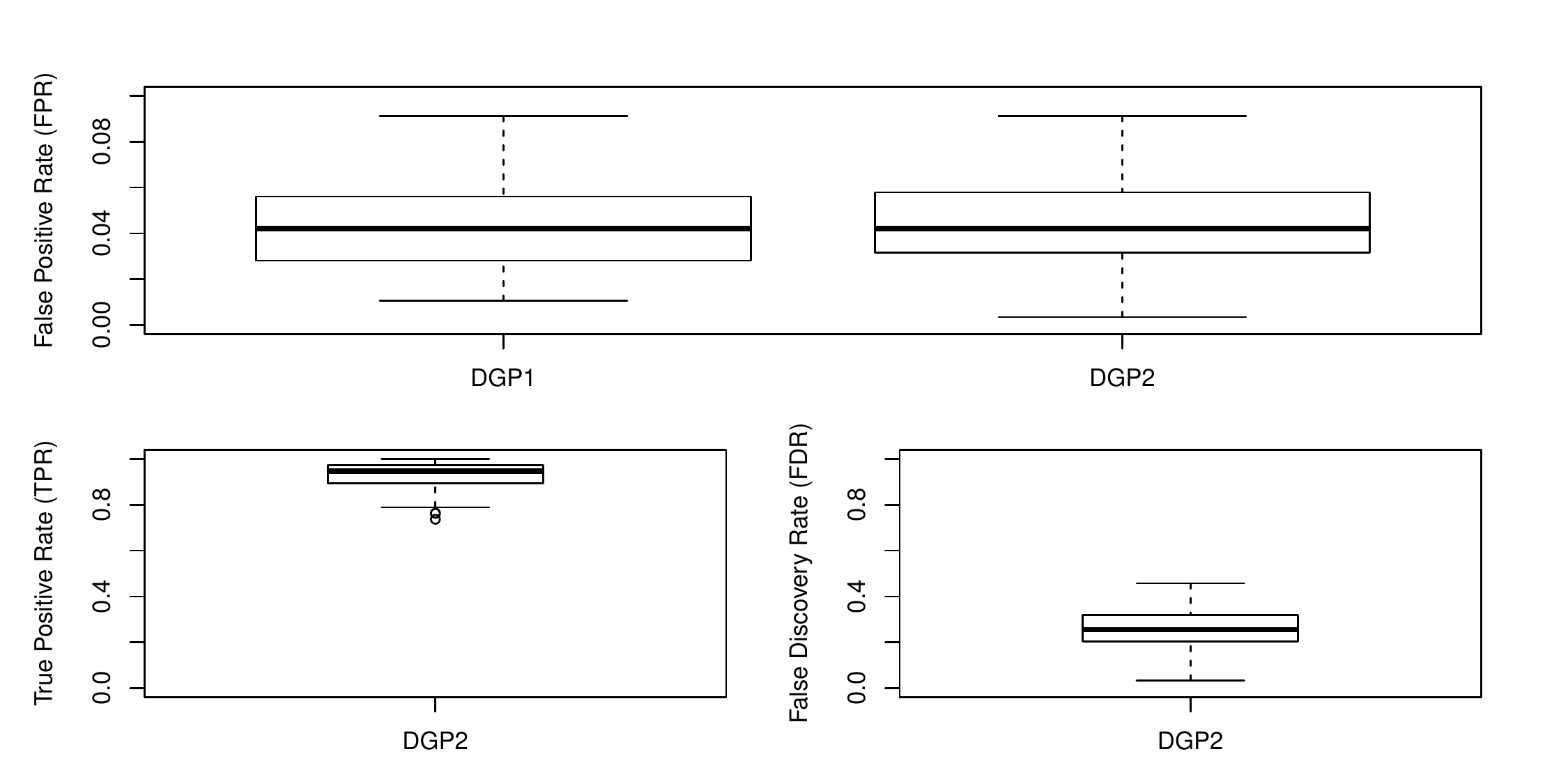}	
\end{figure}

In order to make the following evaluation transparent, we represent the evaluation scheme (e.g. \citealp{fawcett2006introduction}) for both DGPs in Table \ref{tab:scheme}. On the left hand side, we regard the simulation without under-reporting (DGP1). In this setting we can investigate the false positive rate (FPR), being the sum of the false positives (FP) relative to the number of all observations that are in fact not under-reported. A low value for this measures means in DGP1 that in a setting without under-reporting a low share is classified as under-reporting. In DGP2, the measure tells us  whether including true under-reporting in the simulation leads to an increase of misclassified under-reporting. The corresponding results are visualized on the top panel of Figure \ref{fig:eval}. The FPR shows a higher variability in DGP2 (right panel) and is slightly higher as compared to DGP1 (left panel). However, the results provide evidence for an overall low FPR in both setting.

Furthermore, DGP2 allows to evaluate the share of under-reported observations that is  identified. This is assessed based on the true positive rate (TPR) and shown on the bottom left panel of Figure \ref{fig:eval}. In fifty percent of all simulation runs we are able to identify at least 95\% of the falsely censored observations and even in the simulation runs with the worst performance, the TPR are does not fall below 74\%.  Additionally, we investigate the False discovery rate (FDR) that relates the observations that are wrongly classified to be under-reporting to the sum of all observations that are classified for under-reporting. A low value for this measure provides evidence, whether the model is able to keep the number of potential over-reporting that are in fact not under-reporting low. The corresponding results are shown in the south-east panel of Figure \ref{fig:eval}. We find a median share of less than 26\% to be classified incorrectly. 


\subsection{"Forensic" analysis of arms trade data}
\begin{figure}[t!]\centering
	\caption{Densities of the under-reporting network $\Omega_t^+$ and the over-reporting network $\Omega_t^-$ over time.}	\label{figure:dens2}
	\includegraphics[trim={0.1cm 0cm 0cm 0cm},clip,width=\textwidth]{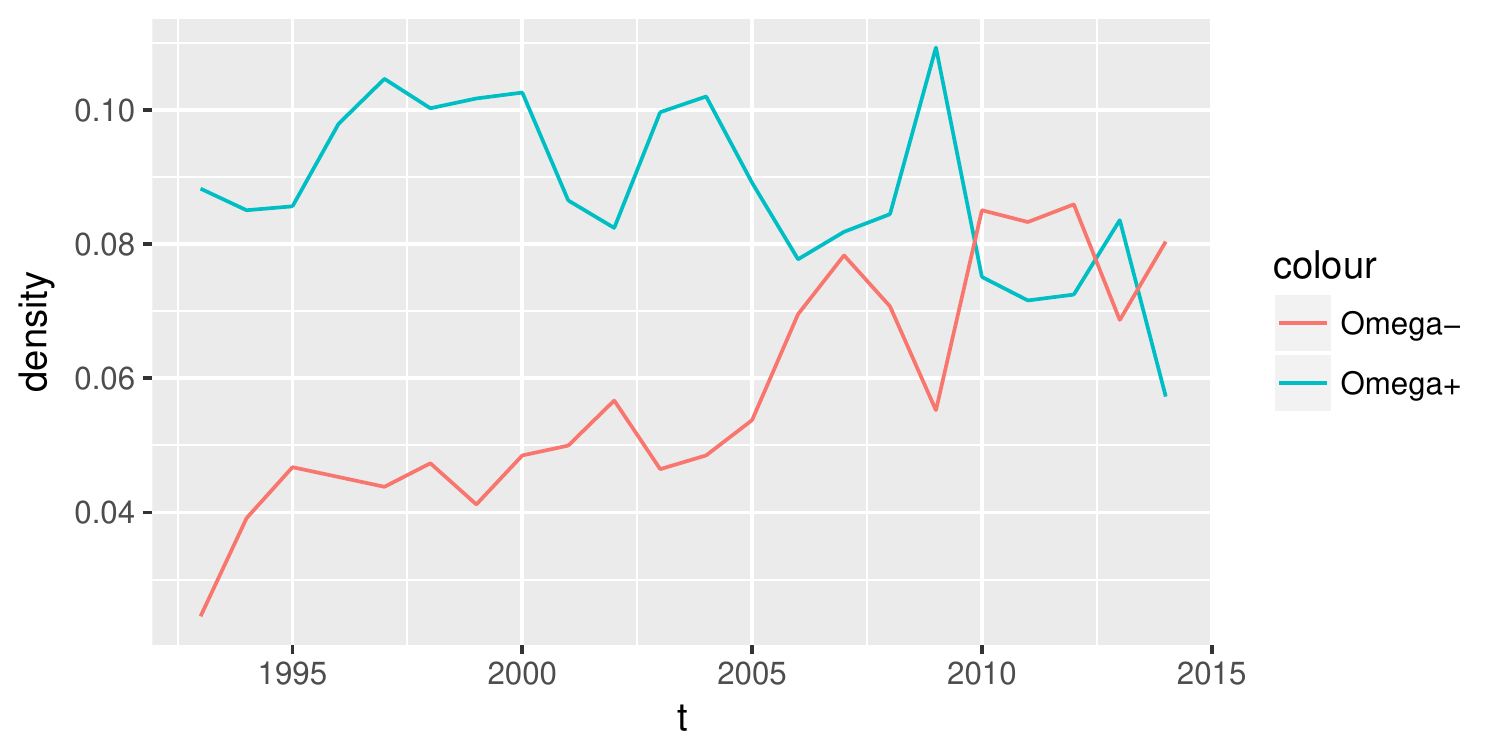}	
\end{figure}

\begin{figure}[t!]
	\caption{Ordered box-plot representation of topological network features of the under-reporting networks $\Omega^+_t$ for $t=1993,...,2014$: Eigenvalue centrality (top), outdegree (middle) and indegree (bottom).}	
	\label{fig:notseen1_1}
	\centering
	\begin{subfigure}{\textwidth}
		\centering\includegraphics[trim={0cm 0cm 0cm 0cm},clip,width=\textwidth]{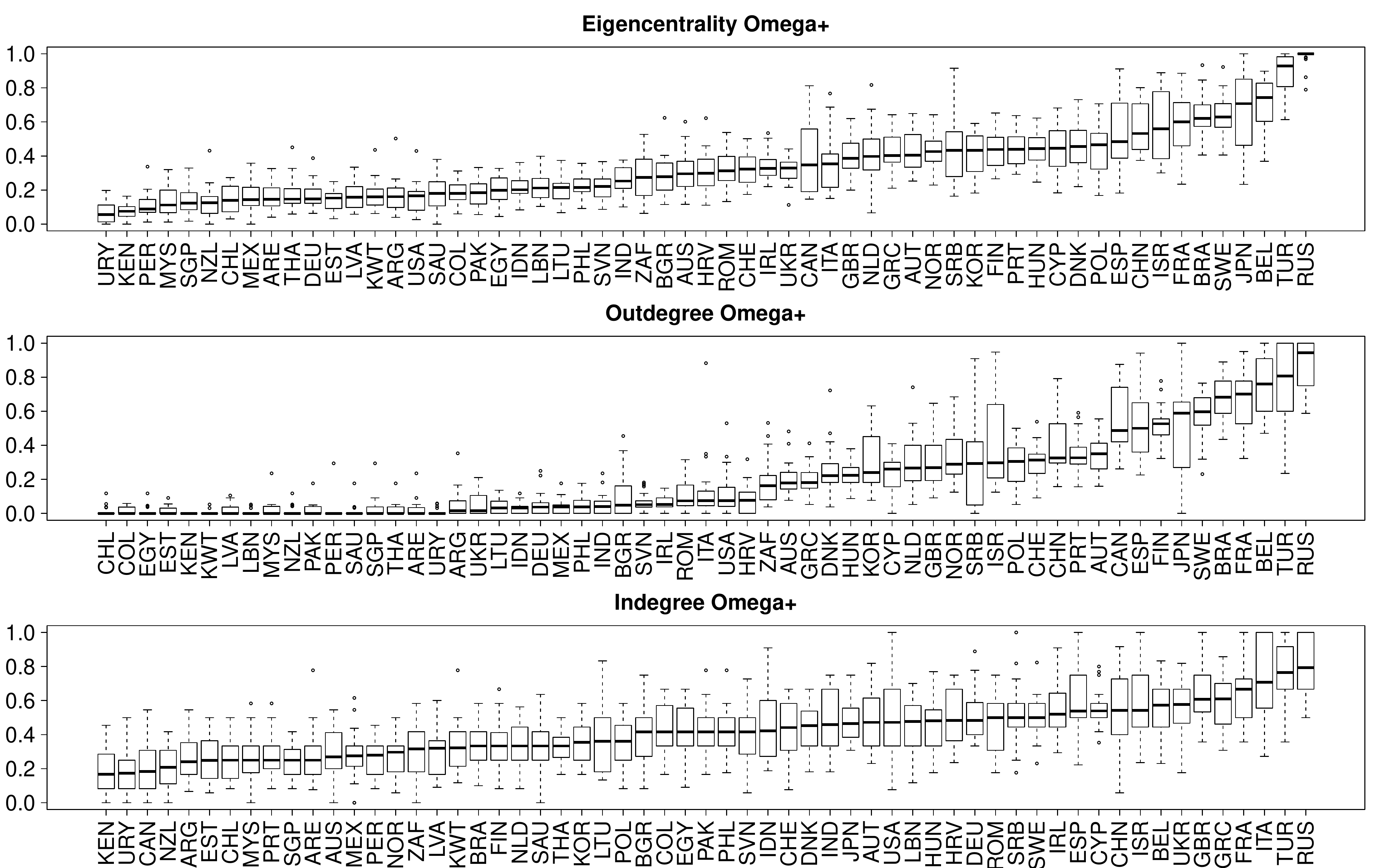}
	\end{subfigure}
\end{figure}
\begin{figure}[t!]
	\caption{Ordered box-plot representation of topological network features of the over-reporting networks $\Omega^-_t$ for $t=1993,...,2014$: Eigenvalue centrality (top), outdegree (middle) and indegree (bottom).}	
	\label{fig:notseen2_1}
	\centering

	\begin{subfigure}{\textwidth}
		\centering\includegraphics[trim={0cm 0cm 0cm 0cm},clip,width=\textwidth]{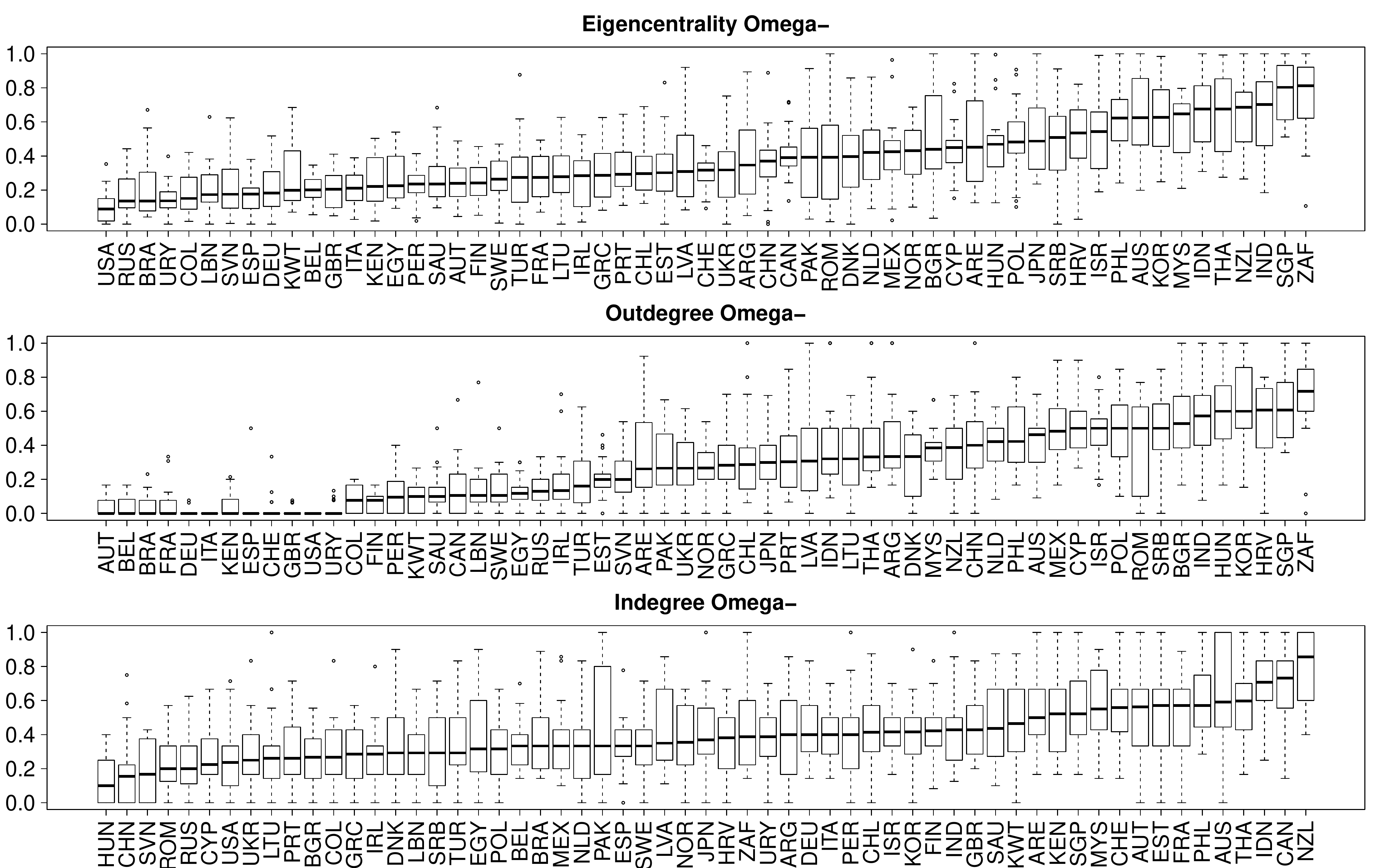}
		
	\end{subfigure}

\end{figure}
We now turn back to the data and provide the development of the densities for the latent networks in Figure \ref{figure:dens2}. In the real data, over- and under-reporting is certainly not random but potentially clustered among countries. We therefore  evaluate node (i.e.\ country) specific network topologies of $\Omega_{t}^+$ and $\Omega_{t}^-$ for each year and summarize the information in box-plots for each country, ordered according to the median of the respective feature. This is shown in Figure \ref{fig:notseen1_1} for potential under-reporting and in Figure \ref{fig:notseen2_1} for over-reporting. In the first row, we represent the Eigenvector centrality scores. This measure is undirected and constructed such that the centrality of each country is proportional to the sum of the centralities of its trading partners. Hence, countries with high scores have many potentially under-reported (over-reported) import- and export-relations with many other countries that themselves have many under-reported (over-reported) import- and export-relations, see e.g. \citet{csardi2006}. In the middle row, we present the outdegree, that is the number of potentially under-reported (over-reported) exports for a country. The bottom row in Figures \ref{fig:notseen1_1} and \ref{fig:notseen2_1} gives the indegree, that is the number of  potentially under-reported (over-reported) imports. All measures are scaled to take values between $0$ and $1$. Countries at the right hand side in the plots of Figure \ref{fig:notseen1_1}  are potentially under-reporting and in Figure \ref{fig:notseen2_1}, the right hand side of the plots mirrors high over-reporting.

\begin{figure}[t!]
	
	\caption{Frequency distribution of transfers in the aggregated under-reporting network ($\Omega^+_{\mathcal{T}}$, black "+") and over-reporting ($\Omega^-_{\mathcal{T}}$, grey "-") networks on the vertical axis. Number of years with under-reporting ($\omega_{\mathcal{T},ij}^+$) or over-reporting ($\omega_{\mathcal{T},ij}^-$) on the horizontal axis. Transfers with the most years predicted are indicated in the form "exporter-importer" in black for $\Omega^+_{\mathcal{T}}$ and in grey for $\Omega^._{\mathcal{T}}$.}
	\centering
	\label{fig:uncover}
	\includegraphics[trim={0cm 0cm 1cm 2cm},clip,width=\textwidth]{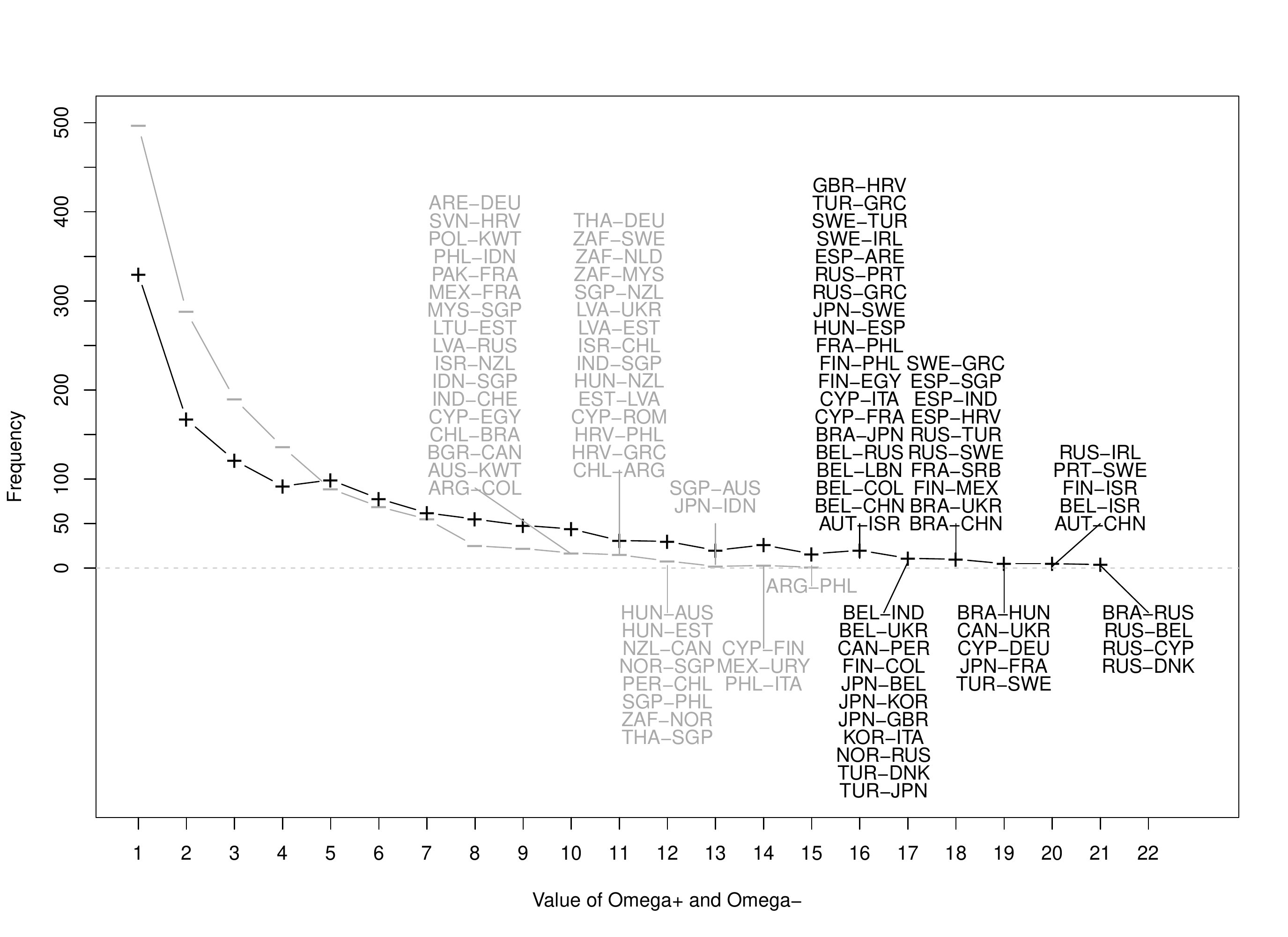}
\end{figure}

To detect persistent patterns in the networks on a dyadic level, we check whether potential under-reporting or over-reporting occurs frequently, i.e.\ counting instances of  $\omega_{t,ij}^+=1$ and $\omega_{t,ij}^-=1$ for $t\in \mathcal{T}=\{1993,...,2014\}$. Denote the aggregated "forensic" networks as
\begin{equation*}
	\begin{split}
		\Omega_\mathcal{T}^+ &=  \sum_{t \in \mathcal{T}} \Omega_t^+ \text{, }\\
		\Omega_\mathcal{T}^- &= \sum_{t \in \mathcal{T}}\Omega_{t}^- .
	\end{split}
\end{equation*} 
We look at the distribution of elements of $\Omega_\mathcal{T}^+$ and $\Omega_\mathcal{T}^-$, which is plotted in Figure \ref{fig:uncover}. On the horizontal axis we show the possible values of the matrix entries, that is
the number years where transfers in the forensic networks occur. This ranges from 1 (potential under-reporting or over-reporting in one year) to 22 (potential under-reporting or over-reporting in all years). The maximum entry of $\Omega_\mathcal{T}^+$ is thereby less then 22, namely 21, while the maximum value of $\Omega_\mathcal{T}^-$ is 15. On the vertical axis of Figure \ref{fig:uncover} we show the frequency of the entries of $\Omega_\mathcal{T}^+$ and $\Omega_\mathcal{T}^-$. Apparently for "forensic" purposes, large values of $\Omega_\mathcal{T}^+$ are of particular interest, since they report pairs of countries which are likely to under-reporting.

The line in solid black with the "+" symbols represents $\Omega^+_\mathcal{T}$ and the line in  grey with the "-" symbols represents the under-reporting network. Additionally, we indicate for both networks the pairs of countries (i.e. sender and receiver) which are of particular interest for "forensic" purposes. This means for example for an element of $\Omega_\mathcal{T}^+$ that has value 21, that the respective transfer from $i$ to $j$  is one of the four transfers appeared that appeared $21$ times in the under-reporting network.

\textsl{Under-reporting networks $\Omega_t^+$:}
Looking at the Eigenvector centrality scores of Figure \ref{fig:notseen1_1} on the top provides conclusive results about countries that are central in the network series $\Omega_{t}^+$. Among the countries where arms transfers are potentially under-reported, we find many Western European countries such as  Belgium (BEL), Sweden (SWE), France (FRA), Spain (ESP) and Denmark (DNK). However, the list of presumed under-reporting is headed by Russia (RUS) and Turkey (TUR) but also Brazil (BRA), Israel (ISR) and China (CHN) have high scores. These countries also play a dominant role in Figure 	\ref{fig:uncover}. In particular, exports from Brazil (BRA) to Russia (RUS), Hungary (HUN), Ukraine (UKR), China (CHN) and Japan (JPN) are likely to be frequently under-reported. Similarly, exports from Russia (RUS) to Cyprus (CYP), Denmark (DNK), Ireland (IRL), Turkey (TUR), Sweden (SWE), Portugal (PRT) and Greece (GRC) are listed. We also find imports of Israel (ISR) from Finland (FIN), Belgium (BEL) and Austria (AUT) as well as exports of Belgium (BEL) to India (IND), Ukraine (UKR), Russia (RUS), Lebanon (LEB), Colombia (COL) and China (CHN).

\textsl{Over-reporting networks $\Omega_t^-$:} 
Among the twelve countries with the highest Eigenvector centrality in Figure \ref{fig:notseen2_1} is Croatia (HRV)  as the only European country. There are however many countries from Asia such as Singapore (SGP), India (IND), Thailand (THA), Indonesia (IDN), Malaysia (MYS), South Korea (KOR) and Philippines (PHL). Furthermore, South Africa (ZAF), New Zealand (NZL), Australia (AUS) and Israel (ISR) are among the countries where trade activity is often over-reported. For Asian countries as well as for Australia and New Zealand this might mirror the fact that those countries export many SAA to Europe and the United States despite the strongly negative distance effect of the model. Furthermore, this network is very likely to be driven by bilateral agreements and historical developments not covered by the covariates. See for example in Figure \ref{fig:uncover} the number of over-reporting related to the Baltic countries Estonia (EST) Lithuania (LTU) and Latvia (LVA).

It remains to be emphasized that the constructions of the forensic networks relies on our model with corresponding assumptions and admittedly high degrees of uncertainty. As a consequence, it does  not allow for definite statements about actual hidden transfers. However, many of the dyads listed in Figure \ref{fig:uncover} indeed have either traded massive amounts of civilian arms (e.g.\ AUT-CHN, BRA-RUS, RUS-BEL, RUS-DNK) or had frequent MCW trade relations (e.g.\ RUS-CYP) but almost no documented small arms transfers for military usage. Additionally, many of the countries that take central positions in the forensic networks are known for not being very transparent with respect to their SAA exports and imports (see e.g.\ the \href{http://www.smallarmssurvey.org/weapons-and-markets/tools/the-transparency-barometer.html}{small arms transparency barometer}). 

\FloatBarrier

\FloatBarrier
\section{Conclusion}\label{sec:conc}
In this paper we have modelled the volumes of international transfers of small arms and ammunition for the years 1992-2014 based on data provided by NISAT. As an analytical tool we combined the gravity model of trade with a modified SAR model that allows to enrich the analysis by endogenous network dependencies, accounting for exporter-related,  importer-related and reciprocal dependency among the transfers in the network. Using a censored normal regression model we are able to include information provided by 
zero-valued transfers. The infeasible likelihood of the censored model is maximized using a Monte Carlo EM algorithm. The fitted model shows strong and stable endogenous network effect, especially related to the sender effect and the receiver effect but also some evidence for reciprocity. Additionally, we find a high coefficient on path dependency and a close connection to the exports of civilian small arms. Conditional on that, the classical gravity hypothesis is confirmed  with respect to the  GDP of the importer and physical distance but only exceptionally with respect to political distance measures and the GDP of the exporter. This contrasts with the MCW network where distance plays no role, where political similarity and GDP of the exporter have a strong impact (see \citealp{Thurner2018}). Actually, this difference is plausible, as the technological requirements for the production of small and ammunition are relatively low, and strategic considerations of world-wide acting countries make geographic distances a negligible factor for MCW trade. 

Building on our latent utility framework we were able to explore latent utility networks. With the construction of under-reporting and over-reporting networks we perform for the first time a forensic approach in this area highlighting especially potentially under-reported exports of Russia and Turkey. We refrain, of course, from making too far-reaching assertions. Note that we do not claim to provide unambiguous claims for intentional false reporting. However, we demonstrate that some zero entries in the SAA trading network tend to be not plausible.

\FloatBarrier

\section*{Acknowledgements}\label{acknowledgements}

We would like to thank Nic Marsh from NISAT for his useful comments and explanations on the data.

\newpage
\bibliographystyle{Chicago}
\bibliography{literature}

\appendix
\newpage
\pagenumbering{Roman}

\section{Annex}
\subsection{Descriptives}
\label{descrannex}
\begin{table}[ht] 
	\caption{Different arms types included in the NISAT dataset with three digit arms category code, weapon type, subcategory and number of transfers in the dataset.
	}
	\label{table:arms_included}
	\centering
	\begin{tabular}{cll}
		\hline
		Code         & PRIO Weapons Type                  &  Subcategories                       \\
		\hline
		200                       & Small Arms                        &                           \\
		210                       &                                    & Pistols \& Revolvers                    \\
		230                       &                                    & Rifles/Shotguns (Military)                \\
		233                       &                                    & Assault Rifles                             \\
		234 &                                    & Carbines                                   \\ 
		235                       &                                    & Sniper Rifles                                 \\
		237                       &                                    & Semi-automatic Rifles (Military)            \\
		239                       &                                    & Shotguns (Military)                        \\
		240                       &                                    & Machine Guns  \\
		243                       &                                    & Sub Machine Guns                          \\
		245                       &                                    & Light Machine Guns                      \\
		247                       &                                    & General Purpose Machine Guns                \\
		250                       &                                    & Military Weapons                            \\
		260                       &                                    & Military Firearms                            \\
		270                       &                                    & Machine Guns All Types                      \\
		300                       & Light Weapons                      &                                           \\
		310                       &                                    & Heavy Machine Guns \textless= 12.7mm       \\
		400                       & Ammunition &                                              \\
		415                       &                                    & Small Arms Ammunition                  \\
		417                       &                                    & Small Calibre Ammunition \textless= 12.7mm    \\
		418                       &                                    & Shotgun Cartridges                                \\
		\hline
		\multicolumn{3}{l}{	Source: \href{http://nisat.prio.org/}{nisat.prio.org}.}
	\end{tabular}

\end{table}

\begin{table}[ht]
	
	\caption{The 59 major exporting and importing countries of the small arms and ammunition dataset with ISO 3 country codes.}
	\label{table:countries_included}
	\centering	\resizebox{\columnwidth}{!}{%
		\begin{tabular}{lclclc}
			\hline
			Country & ISO3 Code &  Country & ISO3 Code &  Country & ISO3 Code\\ 
			\hline
			Argentina & ARG & India & IND & Poland & POL \\ 
			Australia & AUS & Indonesia & IDN & Portugal & PRT \\ 
			Austria & AUT & Ireland & IRL & Romania & ROM \\ 
			Belgium & BEL & Israel & ISR & Russia & RUS \\ 
			Brazil & BRA & Italy & ITA & Saudi Arabia & SAU \\ 
			Bulgaria & BGR & Japan & JPN & Serbia & SRB \\ 
			Canada & CAN & Kenya & KEN & Singapore & SGP \\ 
			Chile & CHL & South Korea & KOR & Slovenia & SVN \\ 
			China & CHN & Kuwait & KWT & South Africa & ZAF \\ 
			Colombia & COL & Latvia & LVA & Spain & ESP \\ 
			Croatia & HRV & Lebanon & LBN & Sweden & SWE \\ 
			Cyprus & CYP & Lithuania & LTU & Switzerland & CHE \\ 
			Denmark & DNK & Malaysia & MYS & Thailand & THA \\ 
			Egypt & EGY & Mexico & MEX & Turkey & TUR \\ 
			Estonia & EST & Netherlands & NLD & Ukraine & UKR \\ 
			Finland & FIN & New Zealand & NZL & Un. Arab Emirates & ARE \\ 
			France & FRA & Norway & NOR & United Kingdom & GBR \\ 
			Germany & DEU & Pakistan & PAK & United States & USA \\ 
			Greece & GRC & Peru & PER & Uruguay & URY \\ 
			Hungary & HUN & Philippines & PHL & -  & -  \\ 
			\hline
		\end{tabular}
	}
	
\end{table}
\FloatBarrier

\clearpage
\newpage

\section{Supplementary Material}
\subsection{Derivatives of the complete log-likelihood}\label{annex:der_complete}
The complete log-likelihood is given by
\begin{equation*}
\ell_{comp}(\theta) =-\frac{N}{2}\log(2\pi \sigma^2)+\log(|A(\boldsymbol{\rho})|)-\frac{(A(\boldsymbol{\rho})\tilde{Y}-X\beta)^\mathrm{T}(A(\boldsymbol{\rho})\tilde{Y}-X\beta)}{2\sigma^2},
\end{equation*}
with score vector
\begin{equation}
\label{eq:complete_ll_score}
\begin{split}
\frac{\partial \ell_{comp}(\theta)}{\partial \beta}=&\frac{1}{\sigma^2}X^\mathrm{T}[A(\boldsymbol{\rho})\tilde{Y}-X\beta]\\
\frac{\partial \ell_{comp}(\theta)}{\partial \sigma^2}=&-\frac{N}{2 \sigma^2} + \frac{\tilde{Y}^\mathrm{T}(A(\boldsymbol{\rho}))^\mathrm{T}A(\boldsymbol{\rho})\tilde{Y}-2\beta^\mathrm{T}X^\mathrm{T}A(\boldsymbol{\rho})\tilde{Y}+\beta^\mathrm{T}X^\mathrm{T}X\beta}{2 \sigma^4}\\
\frac{\partial \ell_{comp}(\theta)}{\partial \rho_k}=&- \text{tr}(B(\boldsymbol{\rho})W_k)\\&-\frac{ \tilde{Y}^\mathrm{T}[-W_k-W_k^\mathrm{T}+2\rho_kW_k^\mathrm{T}W_k+\sum_{l\neq k} \rho_l (W_k^\mathrm{T}W_l + W_l^\mathrm{T}W_k)]\tilde{Y}+2\beta^\mathrm{T}X^\mathrm{T}W_k\tilde{Y}}{2\sigma^2}.
\end{split}
\end{equation}
And the corresponding Hessian results in 
\begin{equation}
\label{eq:complete_ll_hessian}
\begin{split}
\frac{\partial^2\ell_{comp}(\theta)}{\partial \beta\partial \beta^\mathrm{T}}&=-\frac{1}{\sigma^2}X^\mathrm{T}X\\
\frac{\partial^2\ell_{comp}(\theta)}{\partial \beta\partial\sigma^2}&=-\frac{1}{\sigma^4}X^\mathrm{T}[A(\boldsymbol{\rho})\tilde{Y}-X\beta]\\
\frac{\partial^2\ell_{comp}(\theta)}{\partial \beta\partial \rho_k}&=-\frac{1}{\sigma^2}X^\mathrm{T}W_k^\mathrm{T}\tilde{Y}\\
\frac{\partial^2 \ell_{comp}(\theta)}{\partial \sigma^2 \partial \sigma^2}&=\frac{N}{2\sigma^4}-\frac{\tilde{Y}^\mathrm{T}(A(\boldsymbol{\rho}))^\mathrm{T}A(\boldsymbol{\rho})\tilde{Y}-2\beta^\mathrm{T}X^\mathrm{T}A(\boldsymbol{\rho})\tilde{Y}+\beta^\mathrm{T}X^\mathrm{T}X\beta}{\sigma^6}\\
\frac{\partial^2  \ell_{comp}(\theta)}{\partial \rho_k \partial \sigma^2}&=\frac{\tilde{Y}^\mathrm{T}[-W_k-W_k^\mathrm{T}+2\rho_kW_k^\mathrm{T}W_k+\sum_{l\neq k} \rho_l (W_k^\mathrm{T}W_l + W_l^\mathrm{T}W_k)]\tilde{Y}+2\beta^\mathrm{T}X^\mathrm{T}W_k\tilde{Y}}{2 \sigma^4}\\
\frac{\partial^2 \ell_{comp}(\theta)}{\partial\rho_k \partial \rho_k}&=-\text{tr}\bigg{(}B(\boldsymbol{\rho})W_kB(\boldsymbol{\rho})W_k\bigg{)}-\frac{\tilde{Y}^\mathrm{T}W_k^\mathrm{T}W_k\tilde{Y}}{\sigma^2}\\
\frac{\partial^2 \ell_{comp}(\theta)}{\partial\rho_k \partial \rho_l}&= -\text{tr}\bigg{(}B(\boldsymbol{\rho})W_lB(\boldsymbol{\rho})W_k\bigg{)}-\frac{\tilde{Y}^\mathrm{T}(W_k^\mathrm{T}W_l + W_l^\mathrm{T}W_k)\tilde{Y}}{2\sigma^2}.
\end{split}
\end{equation}
Where we use Jacobi's formula (see \citealp{magnus1988}) that allows to express the derivative of a matrix determinant  in terms of the derivative of the matrix and its adjugate ($\text{adj}(\cdot)$). Resulting in 
\begin{equation*}
\frac{\partial \log(|A(\boldsymbol{\rho})|)}{\partial \rho_k} =-|A(\boldsymbol{\rho})|^{-1}\text{tr}[\text{adj}(A(\boldsymbol{\rho}))W_k]=-\text{tr}(B(\boldsymbol{\rho})W_k),
\end{equation*}
for the third equation in (\ref{eq:complete_ll_score}). The differentiation of the trace
\begin{equation*}
\frac{\partial \text{tr}(B(\boldsymbol{\rho})W_k)}{\partial \rho_l} = \text{tr}\bigg{(}\frac{\partial B(\boldsymbol{\rho})}{\partial \rho_l}W_k\bigg{)}=-\text{tr}\bigg{(}B(\boldsymbol{\rho})\frac{\partial A(\boldsymbol{\rho})}{\partial \rho_l}B(\boldsymbol{\rho})W_k\bigg{)}=\text{tr}(B(\boldsymbol{\rho})W_lB(\boldsymbol{\rho})W_k)
\end{equation*}
is used for the sixth and seventh equation in (\ref{eq:complete_ll_hessian}).

\subsection{Practical Implementation of the Algorithm}\label{annex:implement}
The gradient
\begin{equation}
\label{eq:profile_gradient}
\frac{\partial \tilde{Q}(\boldsymbol{\rho}|\theta_0)}{\partial \rho_k }=-\text{tr}(B(\boldsymbol{\rho})W_k)-\frac{N}{2}\frac{R_k^*(\boldsymbol{\rho})-\tilde{Y}^{*\mathrm{T}}H_k(\boldsymbol{\rho})\tilde{Y}^*}{S^*(\boldsymbol{\rho}) - \tilde{Y}^{*\mathrm{T}}(A(\boldsymbol{\rho}))^\mathrm{T}H A(\boldsymbol{\rho})\tilde{Y}^*}.
\end{equation}
can be used to maximize 
\begin{equation}
\label{eq:profile}
\tilde{Q}(\boldsymbol{\rho}|\theta_0)=\kappa + \log(|A(\boldsymbol{\rho})|)-\frac{N}{2}\log\bigg{(}S^*(\boldsymbol{\rho}) - \tilde{Y}^{*\mathrm{T}}(A(\boldsymbol{\rho}))^\mathrm{T}H A(\boldsymbol{\rho})\tilde{Y}^*\bigg{)}.
\end{equation}
by applying the BFGS optimization  routine (see \citealp{broyden1970}, \citealp{fletcher1970}, \citealp{goldfarb1970} and \citealp{shanno1970}). The implementation of the BFGS algorithm in \texttt{R} (\citealp{Rcore2016}) is provided by the base function \texttt{optim}. 
More computational stability for the maximization of equation (\ref{eq:profile}) is reached by defining $\lambda=(\lambda_1,..., \lambda_N)^\mathrm{T}$ as the vector of eigenvalues of $A(\boldsymbol{\rho})$ and replacing  $\log(|A(\boldsymbol{\rho})|)$ by $\sum_{r=1}^{N}\log(\lambda_r(\boldsymbol{\rho}))$ in equation (\ref{eq:profile}), see \citet{bivand2015}.
The starting value for the algorithm can be found by using a maximum pseudolikelihood estimate (MPLE), using $W_1Y,...,W_qY$ as exogenous covariates in a censored regression model, provided by the \texttt{R} package \texttt{censReg} (\citealp{Hennigsen2013}). Since the observed log-likelihood cannot be evaluated, we define $\hat{\theta}$ as the solution of the maximization problem if $(\hat{\theta}-\theta_0)^\mathrm{T}(\hat{\theta}-\theta_0)<0.1$, otherwise we set $\theta_0=\hat{\theta}$ and re-iterate until the stopping criteria is satisfied.

\subsection{Approximation of the Fisher Information}\label{annex:approx:FI}
\citet{louis1982} and \cite{oakes1999} provide formulas for the Fisher information of the observed likelihood. We follow the recommendation of \citet{mclachlan2007}, arguing that Louis's formula is best suited for the MCEM and provides a conservative measure  of the standard errors.
Therefore, we calculate the observed information based on
\begin{equation}
\begin{split}
\label{eq:louis}
-\frac{\partial^2 \ell_{obs} (\theta)}{\partial \theta \partial \theta^\mathrm{T}}&=\mathbb{E}_{\theta}\bigg{[}-\frac{\partial^2 \ell_{comp} (\theta)}{\partial \theta \partial \theta^\mathrm{T}}\bigg{|} \tilde{Y}_o,X,\mathcal{M}\bigg{]}- \mathbb{E}_{\theta}\bigg{[} \frac{\partial \ell_{comp} (\theta)}{\partial \theta}  \bigg{(}\frac{\partial \ell_{comp} (\theta)}{\partial \theta} \bigg{)}^\mathrm{T}  \bigg{|} Y_o,X,\mathcal{M}\bigg{]}\\&+  \mathbb{E}_{\theta}\bigg{[} \frac{\partial \ell_{comp} (\theta)}{\partial \theta}\bigg{|} Y_o,X,\mathcal{M}\bigg{]}\bigg{(} \mathbb{E}_{\theta}\bigg{[} \frac{\partial \ell_{comp} (\theta)}{\partial \theta}\bigg{|} Y_o,X,\mathcal{M}\bigg{]} \bigg{)}^\mathrm{T} .
\end{split}
\end{equation}
Note that the second term of (\ref{eq:louis}) depends not only on the first and second but also on the third and fourth conditional moment of the truncated multivariate normal and cannot be evaluated analytically therefore.

In order to approximate the observed information we are using the results of \citet[p. 187]{robert2004} and \citet[Section 3.7]{kang2013} that allow for an approximation of the observed information based on the score and Hessian of the complete likelihood. Hence, we can use the results from Section \ref{annex:der_complete} for the following procedure.

We draw $w=1\,000$ times potential realizations $\tilde{Y}_{s,sim}$ from the truncated version of 
\begin{equation}
\label{cdf}
\tilde{Y}^c \sim \mathcal{N}_{N_m}(\mu_m+\Sigma_{mo}\Sigma_{oo}^{-1}(\tilde{Y}_o-\mu_o), \Sigma_{mm}-\Sigma_{mo}\Sigma_{oo}^{-1}\Sigma_{om})
\end{equation}
using the package \texttt{TruncatedNormal} (\citealp{botev2017}). Those are stored for each draw $s$ in a vector $\tilde{Y}^*_{s,sim}=(\tilde{Y}_o,\tilde{Y}_{s,sim})$.

Then we calculate the score and Hessian from equations (\ref{eq:complete_ll_score}) and  (\ref{eq:complete_ll_hessian})  $w$ times, where we replace $\tilde{Y}$ by $\tilde{Y}^*_{s,sim}$ in each equation and index them by $s$, allowing to calculate the empirical version of (\ref{eq:louis}) by approximating the expectations by means. 
\begin{equation*}
\begin{split}
&	-\frac{\partial^2 \ell_{obs} (\theta)}{\partial \theta \partial \theta^\mathrm{T}}\approx\\&	\frac{1}{w}\sum_{s=1}^{w}\bigg{[} -\frac{\partial^2\ell_{s,comp}(\theta)}{\partial \theta\partial \theta^\mathrm{T}}-\bigg{(}     \frac{\partial\ell_{s,comp}(\theta)}{\partial \theta} -\frac{1}{w}\sum_{s=1}^{w}   \frac{\partial\ell_{s,comp}(\theta)}{\partial \theta}\bigg{)}   \bigg{(}     \frac{\partial\ell_{s,comp}(\theta)}{\partial \theta}  -\frac{1}{w}\sum_{s=1}^{w}   \frac{\partial\ell_{s,comp}(\theta)}{\partial \theta} \bigg{)}^\mathrm{T}\bigg{]}.
\end{split}
\end{equation*}
This gives an estimator for the observed information. Standard errors are obtained by the square root of the diagonal elements of the inverted approximated matrix. 

\subsection{Data Transformation}\label{annex:transform}
\begin{figure}[t!]\centering
	\caption{Kernel Density Estimate of the log-transformed standardized observed response variable pooled for all time periods (left). Q-Q plot for the log-transformed standardized response variable (right).}	\label{fig:norm_response}
	\includegraphics[trim={0.1cm 0cm 0cm 0cm},clip,width=\textwidth]{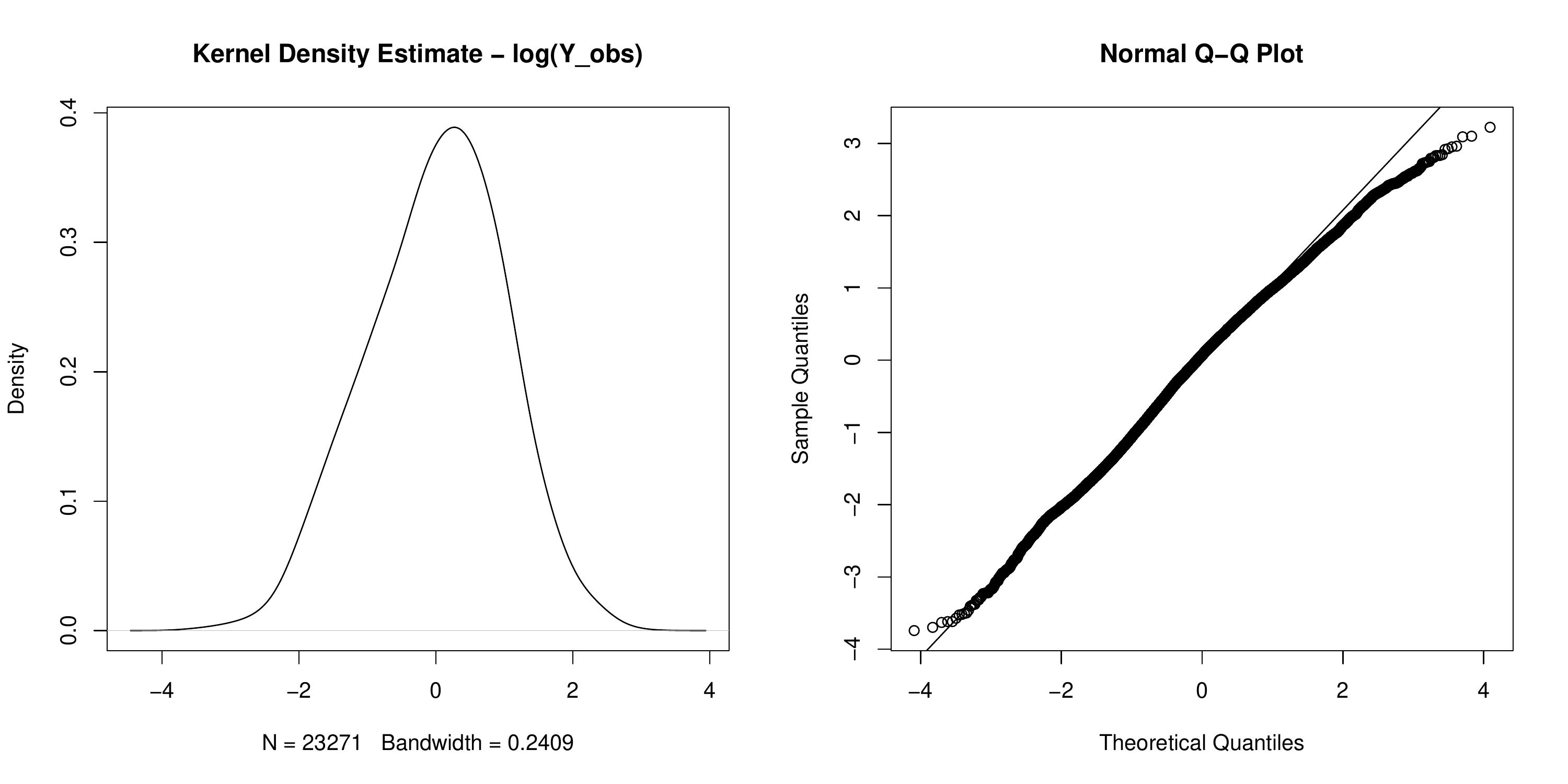}	
\end{figure}

In Figure \ref{fig:norm_response} we show the distribution of the observed log-transformed response variable. The data is pooled over all years and standardized to have mean zero and variance one. The panel on the left side shows a kernel density estimate and the panel on the right gives a Q-Q plot. 

\subsection{Conditional Probabilities}\label{annex:SI}

Based on the fitted coefficients $\hat{\theta}_t=(\hat{\beta}_t,\hat{\boldsymbol{\rho}}_t,\hat{\sigma}_t^2)$ and our model assumptions, we can represent the joint distribution of the latent utility network $Z_t$ via a multivariate normal
\begin{equation*}
Z_t \sim \mathcal{N}_N(\hat{\mu}_t,\hat{\Sigma}_t),
\end{equation*}
where $\hat{\mu}_t=B(\hat{\boldsymbol{\rho}_t})X\hat{\beta_t}$ ands $\hat{\Sigma}_t=B(\hat{\boldsymbol{\rho}_t})(B(\hat{\boldsymbol{\rho}_t}))^T\hat{\sigma}^2_t$.  Given that, define $Z_{t,-ij}$ as the $(N-1)$-dimensional vector, containing all entries of $Z_t$ except $Z_{t,ij}$. Additionally, for example in the case that $ij$ is the first entry of $Z_t$, rearrange $\hat{\Sigma}_t$ such that
\begin{equation*}
\hat{\Sigma}_t=\left(
\begin{array}{cc}
\hat{\Sigma}_{t,ij,ij}&\hat{\Sigma}_{t,ij,-ij}\\
\hat{\Sigma}_{t,-ij,ij}&\hat{\Sigma}_{t,-ij,-ij}\\
\end{array}
\right).
\end{equation*}
Then, the conditional distribution of $Z_{t,ij}$ is given by a univariate normal distribution
\begin{equation*}
\begin{split}
Z_{t,ij}|X,Z_{t,-ij}&\sim \mathcal{N}(\hat{\mu}_{t,ij|-ij},\hat{\Sigma}_{t,ij|-ij} )\text{, where}\\
\hat{\mu}_{t,ij|-ij}&=\hat{\mu}_{t,ij}+\hat{\Sigma}_{t,ij, -ij}\hat{\Sigma}^{-1}_{t,-ij, -ij}(Z_{t,-ij}-\hat{\mu}_{t,-ij})\text{ and}\\
\hat{\Sigma}_{t,ij|-ij}&=\hat{\Sigma}_{t,ij,ij}-\hat{\Sigma}_{t,ij,-ij}\hat{\Sigma}^{-1}_{t,-ij,-ij}\hat{\Sigma}_{t,-ij,ij}.
\end{split}
\end{equation*}
We are interested in a possible state of the network, where the latent utility is allowed to be greater $c_t$. Therefore,  we insert the expectation for the non-observed utility in $Z_{t,-ij}$ and denote this by $\tilde{Z}_{t,-ij}$. Consequently, we can calculate the probability of $Z_{t,ij}$ being greater than $c_t$ using
\begin{equation*}
\begin{split}
\pi_{t,ij} &= \mathbb{P}(Z_{t,ij}>c_t|X_{t,ij},\tilde{Z}_{t,-ij};\hat{\theta})=1-\mathbb{P}(Z_{t,ij}\leq c_t|X_{t,ij},\tilde{Z}_{t,-ij};\hat{\theta})\\
&=1-\int_{-\infty}^{c_t}\frac{1}{\sqrt{2\pi \hat{\Sigma}^2_{t,ij|-ij}}}\exp\left(-\frac{(U-\hat{\mu}_{t,ij|-ij})^2}{2 \hat{\Sigma}^2_{t,ij|-ij}}\right)dU.
\end{split}
\end{equation*}
The probability $\pi_{t,ij}$ can be interpreted as the probability that the latent utility of a transfer from country $i$ to country $j$ is higher than the threshold $c_t$ conditional on the covariates $X_t$ and the remaining network, where no transfer is restricted to be smaller $c_t$.

\subsection{Simulation study - Endogenous effects and approximation of censored variables}\label{sec:sim}
\FloatBarrier
\begin{figure}[t]\centering
	\caption{Results of DGP1. Expected values against true censored values for all simulations (left). Angle bisector in dashed black, non-parametric mean in black and colored contours. Boxplots for the difference between estimated and true values for $\bm{\rho}$ (right).}	\label{figure:dgp1}
	\includegraphics[trim={0.15cm 0.6cm 0cm 2cm},clip,width=\textwidth]{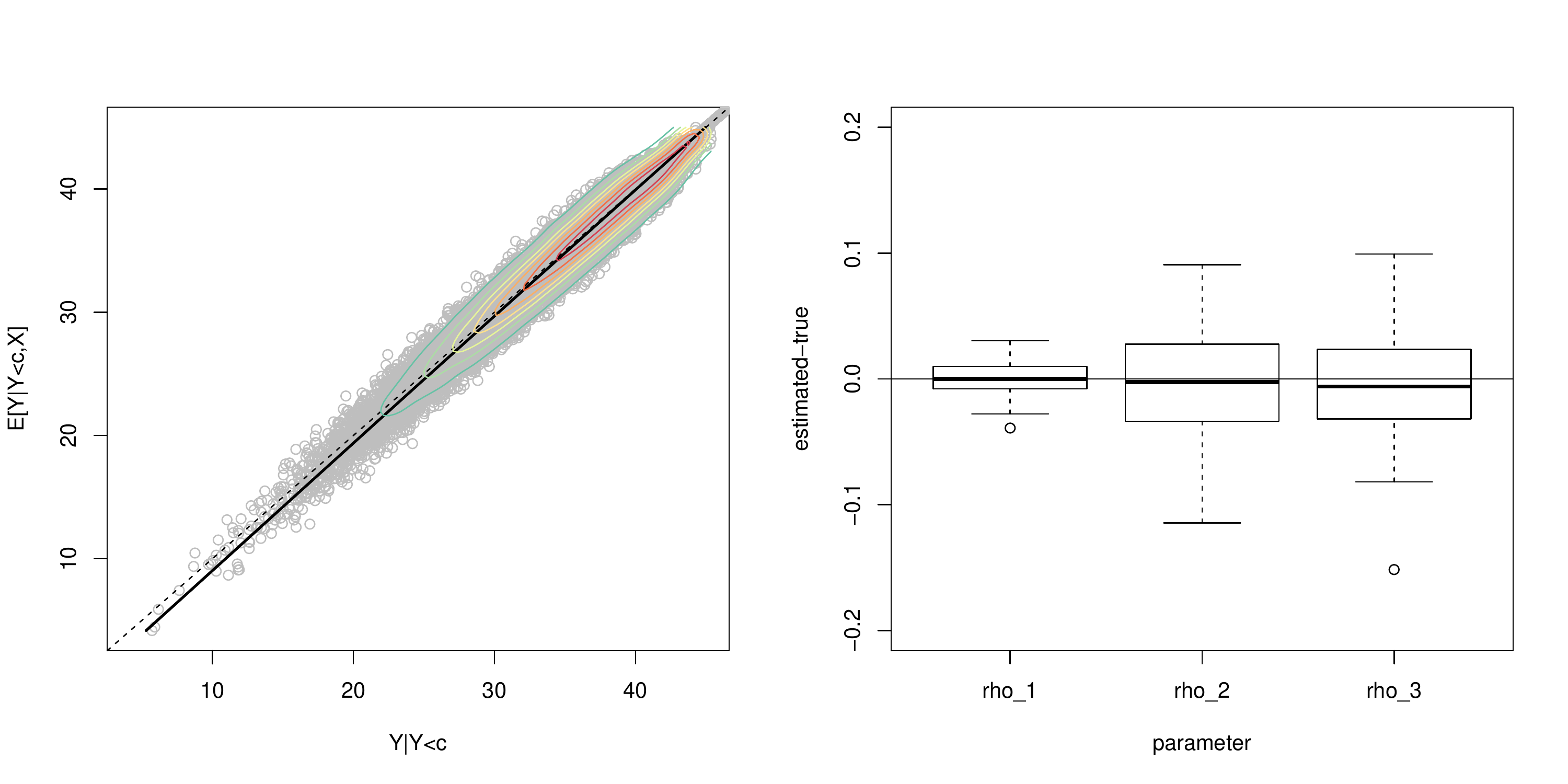}	
\end{figure}
In order to analyse the properties of our estimator, we use the following Data Generating Process (DGP1)
\begin{equation}\begin{split}
\bm{\rho}&=(0.1,0.2,0.3)^\mathrm{T}\text{, }  \beta=(1,2,3,4,5)^\mathrm{T}\text{, }p=5\text{, }n=20\text{, }N=380\\
X&\sim \mathcal{N}_{p}(\bm{1},I_{p})\\
Z&\sim \mathcal{N}_N(B(\bm{\rho})X\beta,B(\bm{\rho})B(\bm{\rho})^\mathrm{T})\\
\tilde{Z}_{ij}&=I(Z_{ij}>q_{0.75}(Z))Z_{ij}\text{, for }i \neq j=1,...,n.
\end{split}
\end{equation}
Here, $q_{0.75}(Z)$ denotes the 75\% quantile and we are censoring the network towards an observed density of $0.25$. Note, that DGP1 is not subject to under-reporting and all censored responses are in fact below the censoring threshold.
The results of running DGP1  $100$ times and applying the estimation procedure are summarized in Figure \ref{figure:dgp1}. On the left panel, we show the true but censored values against the expected values from the last E-Step, together with contour curves and a non-parametric fit for the mean in solid black. It can be seen that the expected values approximate the latent variables very well.  The right panel of Figure \ref{figure:dgp1} shows boxplots for the difference between the true values of $\bm{\rho}$ and the estimated parameters. It indicates that as we are able to find unbiased estimates of the endogenous parameters despite the enormous amount of censoring.

\end{document}